\documentclass{article}

\usepackage{svg}
\usepackage{amsthm}
\usepackage{mathtools}
\usepackage{amsfonts,amssymb,amsbsy}
\usepackage{graphicx}
\usepackage{anysize}
\usepackage{amsmath}   
\usepackage{subcaption}
\usepackage{algorithm}
\usepackage{algorithmic}
\usepackage{ulem}
\usepackage{comment}
\usepackage[utf8]{inputenc} 
\usepackage[T1]{fontenc}    
\usepackage{url}         
\usepackage{booktabs}      
\usepackage{amsfonts}      
\usepackage{nicefrac}       
\usepackage{microtype}      
\usepackage{lipsum}	
\usepackage[authoryear]{natbib}
\usepackage{caption}
\usepackage{subcaption}
\usepackage{amsmath}
\usepackage{tikz}
\usetikzlibrary{shapes, calc, shapes, arrows, snakes}
\newcommand{\width}{0.2}
\usepackage{stfloats}
\usepackage{lipsum}
\usepackage{multicol}
\usepackage{graphicx}
\usepackage{setspace}
\usepackage{PRIMEarxiv}
\usepackage[utf8]{inputenc} 
\usepackage[T1]{fontenc}    
\usepackage{hyperref}       
\usepackage{fancyhdr}       
\graphicspath{{media/}}     

\begin{document}
\begin{center}

{\huge	{\centering \bf{
\textbf\newline{Informative regularization for a multi-layer perceptron RR Lyrae classifier under data shift }}}}
\vspace{3pt}

{\large {\centering \bf{Francisco Pérez-Galarce$^{1}$\footnote[1]{Corresponding author email: fjperez10@uc.cl} , Karim Pichara$^{1,2}$ , Pablo Huijse$^{3,2}$,\\ \centering Márcio Catelan$^{4,2,5}$, 
Domingo Mery$^{1}$}}}
\vspace{2pt}
\\
$^1$Department of Computer Science, School of Engineering, Pontificia Universidad Cat\'olica de Chile. Av. Vicu\~{n}a Mackenna 4860, 7820436, Macul, Santiago, Chile.\\
$^2$Millennium Institute of Astrophysics, Nuncio Monseñor Sotero Sanz 100, Of. 104, Providencia, Santiago, Chile\\
$^3$Instituto de Informática, Facultad de Ciencias de la Ingeniería, Universidad Austral de Chile, General Lagos 2086, Valdivia, Chile\\
$^4$Instituto de Astrof\'isica, Facultad de F\'isica, Pontificia Universidad Cat\'olica de Chile, Av. Vicu\~{n}a Mackenna 4860, 7820436, Macul, Santiago, Chile\\
$^5$Centro de Astroingenier{\'{\i}}a, Pontificia Universidad Cat{\'{o}}lica de Chile, Av. Vicu\~{n}a Mackenna 4860, 7820436 Macul, Santiago, Chile
\end{center}
\vspace{2pt}
\begin{abstract}
In recent decades, machine learning has provided valuable models and algorithms for processing and extracting knowledge from time-series surveys. Different classifiers have been proposed and performed to an excellent standard. Nevertheless, few papers have tackled the data shift problem in labeled training sets, which occurs when there is a mismatch between the data distribution in the training set and the testing set. This drawback can damage the prediction performance in unseen data. Consequently, we propose a scalable and easily adaptable approach based on an informative regularization and an ad-hoc training procedure to mitigate the shift problem during the training of a multi-layer perceptron for RR Lyrae classification. We collect ranges for characteristic features to construct a symbolic representation of prior knowledge, which was used to model the informative regularizer component. Simultaneously, we design a two-step back-propagation algorithm to integrate this knowledge into the neural network, whereby one step is applied in each epoch to minimize classification error, while another is applied to ensure regularization. Our algorithm defines a subset of parameters (a mask) for each loss function. This approach handles the forgetting effect, which stems from a trade-off between these loss functions (learning from data versus learning expert knowledge) during training. Experiments were conducted using recently proposed shifted benchmark sets for RR Lyrae stars, outperforming baseline models by up to 3\% through a more reliable classifier. Our method provides a new path to incorporate knowledge from characteristic features into artificial neural networks to manage the underlying data shift problem.
\end{abstract}


\section{Introduction}
\label{introduction}

In recent decades, there have been several applications of machine learning models in variable star studies, see, for example, \citep{debosscher2007automated,debosscher2009automated,richards2011machine,pichara2012improved,mackenzie2016clustering,aguirre2018deep,naul2018recurrent,becker2020scalable, perez2021informative,COKINA2021100488, zhang2021classification}. The identification of the variable star class given its variability pattern is a very engaging  and challenging problem. The challenge is due to the distinct sources and patterns of variability, which may be exogenous (e.g., planetary systems) or endogenous (e.g., eruptions). Additionally, certain characteristic patterns can be overlapped among the more than 110 classes and subclasses of variable stars \citep{samus2017general}. Overcoming this challenge is crucial for astronomers given the significance of variable stars. Some of them (e.g., RR Lyrae and Cepheids) can be used to calculate reliable distance estimations \citep{beaton2016carnegie}. Moreover, they also have given valuable information about the chemical composition (e.g., helium and heavy elements) within the Local Group of galaxies \citep{pietrukowicz2020properties,dekany2022photometric}.

The amount of data generated by modern telescopes has encouraged the use of complex models to automate certain classification tasks. Examples of relevant machine learning models proposed for variable stars classification include support vector machine (SVM) \citep{debosscher2009automated, benavente2017automatic}, random forest (RF) \citep{richards2011active, bloom2012automating, pichara2012improved, benavente2017automatic, narayan2018machine}, convolutional neural networks \citep{aguirre2018deep}, recurrent neural networks \citep{becker2020scalable},  autoencoders \citep{naul2018recurrent}, and even ad-hoc neural networks for periodic time-series have been proposed \citep{zhang2021classification}. 

Despite this noteworthy progress, the astronomy community has identified a latent drawback in the traditional training and model selection schemes applied to variable star classifiers, i.e., the hold-out and cross-validation methods \citep{debosscher2009automated,richards2011active,masci2014automated,perez2021informative}. These schemes assume the same data distribution for the training and testing sets. However,  light-curves in training sets, i.e. labeled data $\mathcal{D}^{S}$ obtained from catalogs, are shifted with respect to the objects in a real observational environment, i.e., objects from population $\mathcal{D}^{P}$; therefore, those models experience a significant performance decay when they are tested beyond training catalogs \citep{Rebbapragada}. This data shift can be associated with differences between the probability distribution of features ($\mathbf{x}$), labels ($y$) or a mixture of them. \color{black}  

\color{black} This problem, also known as {\em data shift problem}, occurs when we assume that the conditions of the training set generation change, e.g., technological setup, with respect to the environment where the model has to be deployed \citep{candela2009dataset}. Some very close problems are the non representative sample issue, i.e., there are similar conditions in training and testing sets but drawbacks in the sampling procedure \citep{clemmensen2022data}; and the imbalanced set problem, i.e., there is a higher proportion of objects belonging to some classes \citep{chawla2004special}.  \color{black} The data shift has been studied extensively by the machine learning community; \citet{candela2009dataset} provide an excellent introduction to the subject with a description of the possible types of data shift, namely, target shift, covariate shift and conditional shift. 

The covariate shift considers $p(\mathbf{x}^S)\neq p(\mathbf{x}^P)$, and $p(y^P|\mathbf{x}^P) = p(y^S|\mathbf{x}^P)$, where $p(\mathbf{x}^S)$ and $p(\mathbf{x}^P)$ are the density probability distributions of features in training and testing sets, respectively\color{black}; and $p(y^P|\mathbf{x}^P) = p(y^S|\mathbf{x}^P)$ are conditional distributions of the label given the features, in training and testing sets, respectively. \color{black} In other words, a covariate shift problem is present when \color{black} when features in the testing set, e.g. period or amplitude, have a different density distribution with respect to the training set, even considering a representative conditional distribution. \color{black} The target shift assumes  $p(y^P)\neq p(y^S)$, but $p(\mathbf{x}^S|y^S) = p(\mathbf{x}^P|y^P)$.   It is to say a target shift involves a mismatch between the target probability in the training set and the testing set. \color{black} The conditional shift is characterized by $p(y^S|\mathbf{x}^S) \neq p(y^P|\mathbf{x}^P)$ and $p(y^P) = p(y^S)$;  this type of data shift is observed when the relationship of one or all the classes change with respect to the features; for example, for the same feature values a star type can have a different probability value in the testing set from the training set. \color{black}  In the majority of scenarios is very complex to divide these biases, and all of them, in conjunction, directly affect model performance, i.e., $p(\mathbf{x}^S,y^S) \neq p(\mathbf{x}^P, y^P)$, it is to say, there is a difference in the joint probability distribution of labels and features between training and testing sets. \color{black}   

 In the case of variable stars, the underlying shift has two main sources: (i) selection bias \citep{cabrera2014systematic}, i.e., expert labellers assign a class when the object is easier to identify, so that difficult objects are less frequently tagged; (ii)  bias related to the data collection process and the range of information, whereby less luminous objects are more difficult to see; \color{black} hence, even when perfect labels are available or can be obtained, the collected objects are not fully representative for objects beyond this observational setup. 
 
 \color{black} Even though this is a well-known problem in variable star classifiers, and its effects are worrying for the application of machine learning models, few papers have proposed scalable alternatives to reduce the impact of the aforementioned biases in model deployment \citep{debosscher2009automated,richards2011active,masci2014automated,perez2021informative}. An intuitive alternative to tackle the data shift problem is to feed the models with high-level human knowledge, which can mitigate the aforementioned biases. However, the scarcity of solutions with which to confront this problem stems from the difficulty of controlling and injecting knowledge into state-of-the-art machine learning models such as artificial neural networks (ANNs). 
 \color{black}
 
Recently, \cite{perez2021informative} proposed an informative Bayesian model selection for RR Lyrae, showing that simple deterministic rules (DRs) can improve the model selection phase when this kind of knowledge is incorporated into the marginal likelihood estimation. These DRs are a simple representation of knowledge collected from astronomical sources. DRs represent ranges for characteristic \color{black} invariant \color{black} features of variable stars, such as period or amplitude. With that in mind, we propose using this simple representation of human knowledge to design a novel regularization strategy on ANN models. For experimental purposes, a multi-layer perceptron (MLP), a feedforward neural network \citep{svozil1997introduction}, was selected for the following reasons. First, simplicity; second, because the expressiveness of an MLP is adequate to validate the methodological approach of this research, even though one hidden layer on its own is capable of obtaining universal approximation \citep{hornik1989multilayer}.

 \color{black}
In this paper, we propose a scalable and easily adaptable regularization scheme to incorporate astronomical knowledge into ANNs. Our approach only requires a set of DRs which are used to generate synthetic training cases; these training cases are used to regularize the optimization process, mitigating the underlying data shift problem of labeled objects. We provide an alternative to train more reliable RR Lyrae binary classifiers,  even in the presence of the data shift problem. We also design an ad-hoc modeling and training procedure to force an effective informative regularization. Our algorithm considers a double error propagation in each epoch, using two disjoint sets of weights (masks). The first propagation reduces the classification error, and the second propagation spreads the informative regularization through the network. The training procedure also considers an initial phase where the weights are assigned to each loss function. \color{black}  

The paper is organized in the following manner: Section \ref{review} provides a summary of the main related work, considering machine learning models for variable stars. Section \ref{background} presents techniques for injecting human knowledge into machine learning models and; in particular, methodological proposals of informative regularization in ANNs are shown. The methodology used for incorporating human knowledge is presented in Section \ref{method}. The shifted data for RR Lyrae classification and the implementation description are given in Sections \ref{data} and \ref{implementation}, respectively. Experiments comparing the approach undertaken with baseline alternatives are outlined in Section \ref{results}. Finally, conclusions and future work are discussed in Section \ref{conclusions}.

\section{Literature review}
\label{review}

\color{black}
 
 \subsection{Traditional variable stars classifiers}
 \label{shallow}
 Several papers have exploited and applied traditional classifiers, such as tree-based models, as well as logistic regressions (LRs), shallow ANNs, support vector machine, Gaussian mixture model classifiers (GMMC), k-nearest neighbor (KNN) and Bayesian networks (BNs) \citep{debosscher2007automated,debosscher2009automated,richards2011machine,pichara2012improved,mackenzie2016clustering,benavente2017automatic,nun2014supervised,nun2015fats,kim2016package,feets}. The feature engineering that supported these applications showed how they can be used to speed up, for example, the detection of peculiar objects \citep{nun2014supervised}; the automatic classification of catalogs \citep{pichara2012improved,nun2014supervised,pichara2016meta}; and the modeling of domain adaptation on different telescopes \citep{benavente2017automatic}. In fact, some specialized computational packages were designed to extract features from light curves \citep{nun2015fats,hartman2016vartools,vanderplas2016gatspy,naul2016cesium,barbary2016sncosmo}.
 
 Due to the fact that the primary objective of these approaches was to extract knowledge from labeled objects, most of them only focused on the model performance within catalogs; consequently, they did not worry about the data shift problem of these labeled objects. A couple of papers contributing to the data shift problem during this phase were \citet{richards2011active}  and  \citet{masci2014automated}. \citet{richards2011active} highlighted the problem and proposed certain strategies (active learning and importance weighted cross-validation) to improve the training data.  \citet{masci2014automated}  also proposed an active learning method to train an RF with a labeled set from the Wide-field Infrared Survey Explorer \citep[WISE;][]{wright2010wide}. Both approaches are based on active learning; which is a human-in-the-loop approach, where an expert provides prioritized object labels for improving the classification performance \citep{settles2009active}. This method can be highly time-consuming and prone to error, as these experts may also add biases to the labeled data. Hence, a complementary approach to face those biases is needed.  
 \color{black}  

\subsection{Variable stars classification with deep learning}
\label{deep}
The availability of more extensive catalogs of light curves, in conjunction with certain breakthroughs in machine learning and artificial intelligence (e.g., deep learning), led to the incorporation of more complex machine learning models for variable stars classification  \citep{narayan2018machine,naul2018recurrent,carrasco2018deep,aguirre2018deep,becker2020scalable,jamal2020neural,zhang2021classification}. Despite this remarkable progress, there is not contributions for mitigating the data shift problem in deep ANNs. \color{black}

\citet{naul2018recurrent} proposed a recurrent autoencoder to improve understanding of variable star embedding. The measurement error in observations was considered by weighting the reconstruction error in the loss function. The embedding was then used for classification purposes with an RF. Results show that the automatic features extracted by the autoenconder are competitive with engineered-features from \citep{kim2016package,richards2011machine} in terms of classification error. 

\citet{carrasco2018deep} proposed a recurrent convolutional neural network for classifying seven classes of astronomical objects directly from images. They train this model using synthetic images and propose a fine-tuning stage. This model performed similarly to a traditional RF based on handcrafted features. \citet{aguirre2018deep} implemented a convolutional neural network using data from multiple catalogs, outperforming an RF trained using handcrafted features. \citet{becker2020scalable} proposed a recurrent neural network using raw light curves, which performed in a similar way to a feature-based RF, albeit in a shorter timeframe. 

Recently, \citet{jamal2020neural} compared a set of neural network architectures. Four well-known layers were compared: long short term memory (LSTM), gated recurrent unit (GRU), temporal convolutional neural networks (tCNN) and dilated tCNN (dtCNN). They studied two types of architectures; (i) a direct classification scheme, where information was encoded, and after that, this representation was used in a classifier; and (ii) a composite scheme, that considers reconstruction and classification simultaneously. The same strategies were applied in a multi-band setting, where two architectures to merge bands were tested (before and after encoding). Early stopping and dropout techniques were implemented to reduce the over-fitting; however, no comments about the data shift problem were given. Some sub-classes (e.g., RR Lyrae sub-classes) were classified erroneously even when metadata were incorporated; they recommended the injection of prior knowledge about the importance of each feature.  

\citet{zhang2021classification} proposed a new ANN architecture called cyclic-permutation invariant neural network. This novel approach was based on the representation of phased light curves through polar coordinates. Experiments using two implementations (temporal convolutional neural network and residual neural network) were conducted, outperforming non-invariant baseline models in three catalogs; namely, the Optical Gravitational Lensing Experiment \citep[OGLE;][]{udalski2008optical}, the MAssive Compact Halo Object \citep[MACHO;][]{alcock1997macho} and the All-Sky Automated Survey for Supernovae \citep[ASAS-SN;][]{jayasinghe2019asas}. A randomized and stratified split of the aforementioned catalogs (training, validation and testing sets) was applied to assess the generalization capability.

To summarize, ANNs are becoming a crucial technology to speed up discoveries in variable stars; hence, the incorporation of new methods to ensure their generalization capability is a must.  

\subsection{Machine learning classifiers for RR Lyrae stars}
\label{rrlyraeclassification}

Due to their relevance, some articles have developed machine learning models for classifying certain classes of variable stars; here, we focus on RR Lyrae classification. 

Based on their pulsation modes, RR Lyrae can be divided into ab-type (RRab), c-type (RRc), double-mode (RRd), and  e-type (RRe). RRab stars pulsate in the fundamental radial mode, RRc stars pulsate in the first-overtone radial mode, RRd stars pulsate simultaneously in the fundamental and first-overtone modes, and RRe stars may pulsate in the second overtone \citep[but see, e.g.,][and references therein]{catelan2014pulsating}.

\citet{gran2016mapping} designed a procedure to search for RR Lyrae stars considering two steps; first, a candidates generation phase which is based on features analysis; and second, automatic classification using machine learning models.  The first step considers a magnitude variability test and filtering based on the characteristic range for the estimated period. The machine learning models were based on feature engineering of light curves, and they were trained to classify RRab stars. This search focused on the Galactic bulge's outer part, as observed by the VISTA Variables in the Vía Láctea (VVV) ESO public survey \citep{minniti2010vista}. More than 1,000 RRab stars were found using this procedure. \color{black}  

\citet{elorrieta2016machine} compared a set of machine learning classifiers for RRab in VVV. These models were trained using 68 features. They conducted two approaches to assess the model performance, cross-validation and a test over two independent sets, achieving similar results. While these results are interesting, the shift level (in features space) between each independent set and the training set was not studied. 

\citet{sesar2017machine} proposed a gradient tree boosting classifier to identify RR Lyrae stars in multi-epoch and asynchronous multi-band photometric data from the Panoramic Survey Telescope and Rapid Response System \citep[PanSTARRS1;][]{kaiser2010ground}. The training set considered 1.9 million objects, and a three-stage classification scheme was applied. The first classifier, which included 20 features, reduced the candidates from 1.5 million objects to 1,500. The second maintained ten features from classifier one and added 40 features from periodograms (periods and their powers), improving by 15.0\% the purity of the extracted RR Lyrae sample, as compared with previous filtered stars. Finally, a multi-class (RRab, RRc, and non-RR Lyrae) classification using 70 features (multi-band and features used in previous stages) was applied to the remaining objects (910, with 95.0\% completeness and 66.0\% purity). They highlighted that this stage was highly time-consuming ($\sim$30 minutes per object); because of this problem, they needed to filter data in the previous steps.   Additionally, for multi-band period estimation, a template fitting procedure was conducted; for this, a physical informed modeling was applied to find the period, and characteristic ranges for each subclass were considered to constrain the period search during the estimation. 

\citet{dekany2020near} designed a bidirectional long-short term memory recurrent neural network to separate RRab stars from other periodic variable stars in the VVV survey \citep{minniti2010vista}. Unlike recent end-to-end deep ANNs approaches for classifying variable stars \citep{aguirre2018deep,becker2020scalable}, they decided to use phased-folded light curves and provide the model: magnitudes, binned phase from period estimation ($P$), binned phase with $2P$, and $P$. The direct incorporation of period estimation had two reasons: an inexpensive computation in the studied survey since this has low number of observations per object; and the relevance of this feature for separating RRab from other variable star types. They applied the same object selection strategy for both sets to minimize the shift between training and testing sets. The approach has a $\sim$99\% precision and recall metrics when objects with a signal-to-noise ratio above 60.           
\color{black} 

Recently, to face the underlying data shift problem in labeled objects, \citet{perez2021informative} proposed a Bayesian approach for selecting RR Lyrae classifiers considering expert knowledge (e.g., period and amplitude); they incorporated simple rules into a bridge sampling method. The method was tested on data from OGLE and outperformed traditional cross-validation even when penalized models were compared. However, this model selection strategy was time-consuming, so its scalability is limited. \color{black}

Despite the substantial progress achieved in the automatic classification of variable stars, previous approaches have not considered a definitive solution for the underlying data shift (training sets versus testing sets), which is a relevant issue in variable star classification. Our paper focuses on mitigating the data shift problem by incorporating human knowledge into the model, particularly during the training phase of an MLP. The following section discusses specific strategies that can be used to inject this valuable knowledge.

\section{Background theory}
\label{background}


\subsection{Prior knowledge injection into neural networks }
\label{humanNN}
The injection of human knowledge in ANNs (and into machine learning models in general) is one of the longest-standing challenges in artificial intelligence. \citet{deng2020integrating}, \citet{von2019informed} and \citet{borghesi2020improving}   provide comprehensive literature reviews about how human knowledge has been integrated into machine learning models. Most of these approaches have used inductive biases, i.e., a prioritization of solutions independent of the observed data \citep[e.g., regularizers, prior distributions, or data augmentation;][]{battaglia2018relational}. The main lines of research undertaken in this area are presented below: 

\textbf{Data augmentation:} Using human knowledge to generate synthetic data is a common strategy for data quality improvement. This data can be generated by using: simulation techniques from analytical models \citep{blanton2007k}; applying simulation based on machine learning model \citep{sravan2020real}; or using domain knowledge to modify the available data. To illustrate, in the domain of images, knowledge about image invariance is the basis for several standard alternatives of data augmentation, such as flipping, cropping, scaling and translation \citep{shorten2019survey}. In the case of light curves, few papers have focused on data augmentation \citep{castro2017uncertain, aguirre2018deep}. In this domain,  some conditions limit these approaches, such as the temporal dependency of observations; often, light curves must be modeled as a multivariate time series (magnitude observed in more than one band); and the time between observations is irregular. Lastly, to avoid more biases due to the augmented data, it is necessary to know the patterns less represented, and in light curves, that can be challenging to detect in advance.\color{black}  
 
\textbf{Bayesian modeling:} This is an intuitive approach for adding human knowledge to machine learning models. The prior distributions offer a direct path for the provision of expert knowledge. However, literature on informative prior distributions for complex models (e.g., MLP) is scarce \citep{hanson2014informative,fortuin2022priors} and obtaining the true posterior is highly time-consuming.

 \textbf{Knowledge base:} Typically represented by graphs, knowledge bases provide a structure that offers a simple alternative for storing knowledge. Examples include Freebase \citep{bollacker2007freebase}, DBpedia \citep{auer2007dbpedia}, and YAGO \citep{suchanek2007yago}. Knowledge graphs consider entities represented by nodes and relations represented by arcs. The knowledge contained therein is declarative as it uses symbolic language and has been used to generate new data and provide explainability.  \citet{kafle2020overview} review how knowledge bases are incorporated into ANNs. In broad terms, there are two well-known drawbacks in this structured knowledge base, and these relate to completeness and compatibility. The completeness issue comes from the fact that no graph contains all the available information. In specific scientific disciplines, such as astronomy, compiling this knowledge base could be hugely challenging. The compatibility problem relates to the design decisions of each graph, whereby each decision may consider different relations among entities.
 
  \textbf{Feature space:} Traditional mechanisms for feature selection are based on compressing original space or selecting an optimal subset of features. Such methods are purely data-driven and prior knowledge is not included therein. \citet{atzmueller2017mixed} propose generating an additional set of features using a domain-specific knowledge graph for combining pairs of features. They suggest that this knowledge graph successfully manages the level of interaction among features during the knowledge-based feature generation.  
 
 \textbf{Hypothesis space:} The hypothesis space  considers the weights and architecture \citep{borghesi2020improving}. Direct knowledge injection on weights can be performed using constraints or penalisations during the optimization process. The knowledge injection on architecture is incorporated using ad-hoc layers. For example, convolutional layers use knowledge about the correlation of neighbor pixels, recurrent layers use temporal relations knowledge, and graph neural networks have been designed to use graph-expressible prior information.        

 \textbf{High-level knowledge:} This approach focuses on incorporating physical or logic rules into machine learning models. The physical rules can be represented by differential equations, while logic rules are supported by first-order logic (FOL). Both knowledge classes can be incorporated as a new loss function. \citet{raissi2019physics} proposed physical informed ANNs to manage the learning process under a small data set. This type of neural network is constrained to certain symmetries, invariances, or conservation principles whose origin comes from the physical laws and can be represented by non-linear partial differential equations. In this setting, two functions are included: first, a loss minimizing the initial and boundary data, and second, a loss in learning the structure imposed by the non-linear partial differential equation. 

Based on these contributions, our paper proposes to emphasize the mitigation of the data shift problem in MLPs by combining: (i) A knowledge representation using knowledge rules via highly informative rules; (ii) sparse training cases generation, \color{black} from now on signals, \color{black} based on this knowledge representation; and (iii) prioritization of solutions (weights) coherently with DRs through regularization and an ad-hoc training procedure.

It should be noted that a significant component of the research undertaken for this paper is the regularization scheme; therefore, this topic will be reviewed in the next section.

\subsection{Regularization strategies}
\label{regul}


\subsubsection{Non-informative regularizers}

\label{noninfo}
Regularization in ANNs includes a family of strategies focused on improving generalization capability, i.e., reducing the over-fitting issue on over-parameterized ANNs.   \citet{goodfellow2016deep} reviews the main techniques within the regularization framework, such as norm penalties, dataset augmentation, noise robustness, early stopping, parameter tying/sharing, sparse representation and  dropout. Each of these methods have been applied intensively in recent years; however, a gap remains concerning the injection of human knowledge in order to mitigate overfitting. It should be noted that overfitting on shifted data is intractable if we only let the data speak since no reliable data sets are available to measure the performance on unseen shifted data. Thus, knowledge incorporation beyond the training data set is critical.

\citet{neyshabur2014search} discussed the inductive bias, which can be understood as a capacity controller for improving generalization. This inductive bias must consider appropriate flexibility to fit the data accurately. A good example of this is the ANN architecture. Indeed,  \citet{neyshabur2014search} used the number of hidden units for experimental purposes by running a comparison with the matrix factorization strategy. They studied the impact on training/testing using a dimensional control (hidden size) versus norm control (weights). The authors comment that their optimization strategy incorporates a regularization bias towards “low complexity” global minima.

\citet{leimkuhler2020constraint} proposed a constrained approach to regularize ANNs, which the authors apply into a stochastic gradient Langevin optimization algorithm. These algebraic constraints control the weight values and, with this in mind, the authors propose the following three constraints: circle constraints, sphere constraints and orthogonality constraints. The generalization capability is improved when constrained training and unconstrained approaches are compared to training MLPs on different image sets for classification tasks. Despite achieving good performance, there is a lack of human knowledge injection using these constraints.  

A general strategy to induce regularization on machine learning models by means of a penalized component can be defined as follows: 
\begin{align}
g(\beta) = f(\beta) + \lambda\Omega(\beta) \label{eq3},
\end{align}
\noindent where $\beta$ represents the model parameters; $f(\beta)$ is a score function, assessing the prediction quality; and $\Omega(\beta)$ is the penalized component, which is a function of $\beta$, typically, a norm. $\lambda$ is a tuneable hyperparameter that controls the relative importance of the penalized component with respect to the score function during the optimization process.

A well-known penalized approach, which is used from simple regression to complex deep neural networks, is the least absolute shrinkage and selection operator  \citep[Lasso or $l_1$-norm;][]{tibshirani1996regression}. This regularization corresponds to a norm-based regularizers. In a classification model using $l_1$-norm, $f(\beta)$ can be a cross entropy function, $\beta$ represents the model parameters that are controlled/penalized by the sum of absolute values of weights $||\beta||_1$, which is represented by $\Omega(\beta)$ in equation (\ref{eq3}). \color{black} A cross-polytope defines the feasible solutions space for the weights from a geometric interpretation. This regularization has a Bayesian interpretation considering a Laplacian prior over the weights. Such a method lacks human knowledge to define the space of feasible solutions and only the relationship between complexity and weight values is used. This relationship comes from the fact that large weight values increase the output variance; then, according to the bias-variance error decomposition, a large variance increases the generalization error \citep{geman1992neural, svozil1997introduction}. \color{black} 

Similarly, ridge regression constrains the weights by the $l_2$-norm. Here, the geometric interpretation is associated with a hyper-sphere, and the Bayesian perspective is linked to a Gaussian prior \citep{ridge}. Unlike Lasso, the ridge model does not provide a straightforward model interpretation because the parameters only reduce their absolute weight values while remaining non-zero. By focusing on interpretability and as a result of the poor performance of Lasso regression when the number of features exceeds the number of observations ($p \gg n$ problem), \citet{zou2005regularization} proposed the elastic net. The latter can be considered a convex combination between the $l_1$-norm and $l_2$-norm. Unlike the $l_1$-norm, this generalization can manage the grouping problem, which occurs when a group of features has a high pairwise correlation. Conversely, the $l_1$-norm selects only one variable and places no importance on which one it is  \citep{zou2005regularization}.    

More recently, \citet{kim2021novel} have proposed a regularization approach to control the well-known imbalance problem. This regularizer considers two components of the output model distribution: a mean divergence regularization, 
 
\begin{align} \Omega(\beta) = \dfrac{1}{2} \left[\dfrac{1}{2}-E_x \left[ f_{\beta}\left(x\right)\right]\right]^2, \end{align}
 
\noindent where $f_{\beta}$ function represents the classifier; and a variance divergence regularization based on KL divergence, 
 
\begin{align} \Omega(\beta) = \dfrac{1}{2} \left[ D_{KL}(Z_+||Z_-) + D_{KL} (Z_-||Z_+)   \right], \end{align}
 
 \noindent where $Z_+$ and $Z_-$ represent positive and negative training cases, whose distributions are assumed to be $\mathcal{N}(\mu_1, \sigma_1)$ and $\mathcal{N}(\mu_2, \sigma_2)$, respectively. 

These regularization methods outperformed baselines models, mitigating the imbalance problem in classifying sentences and images. They also highlighted the challenge posed by the selection of the $\lambda$ hyperparameter. Hence, new methods avoiding this hyperparameter search are crucial.        
\color{black}

To summarize, several ideas have been proposed to regularize weight-based models, including sparseness promoting, weight controlling, and ad-hoc loss functions for the imbalance problem. However, these approaches have been focused on improving the generalization of models trained using representative data, therefore, when the data are shifted these approaches become futile. Therefore, combining such methods with human knowledge is necessary to improve the generalization capability in this complex setting.   

\subsubsection{Informative regularizers}
\label{info}
Only a limited number of papers have incorporated knowledge into regularization components, and this section provides examples of interesting approaches in that direction. 

The application of the aforementioned noninformative approaches as regularizers would lead to an unstable model (i.e., one that is vulnerable to resampling). To face this drawback, \citet{baldassarre2012structure} compared the $l_1$-norm, elastic net, sparse Laplacian and total variation methods with certain alternatives that exploit spatial information. \citet{chambolle2004algorithm} proposed the total variation method, which uses an additional component to the penalization term: 

\begin{align}  \Omega(\beta) = ||\nabla \beta||_1 + ||\beta||_1,   \end{align}

\noindent where $\nabla \beta$ represents the difference between  neighbors (e.g., voxels or pixels). For example, the difference in the first dimension can be computed as  $(\nabla \beta)_{ijk}^1 = \beta_{ijk} - \beta_{(i+1)jk}$. This regularization on the weights promotes similar behavior in neighboring voxels, in addition to sparsity. Sparse Laplacian offers a soft alternative in which the constant requirements are relaxed by the incorporation of an additional tunable parameter $\alpha$,
\begin{align} \Omega(\beta) =  (1- \alpha)\sum_{(i,j)\in\xi_G }(\beta_i-\beta_j)^2  +  \alpha||\beta||_1,   \end{align}
with $\xi_G$ being the set of edges over the graph of connections $G$. If $\lambda_1 = \alpha \lambda$ and $\lambda_G = \lambda (1-\alpha)$, sparse Laplacian is equivalent to graph Laplacian elastic net (graph-net). On the basis of the spatial informative regularizer,  \citet{jenatton2012multiscale} proposed a hierarchical and spatial informative penalization. In this approach, the neighborhood $\xi_G$ can be represented by using a tree, which is defined from a spatially constrained agglomerative clustering. This spatial and hierarchical regularizer provides enhanced interpretability and improves accuracy in predictions. This type of spatial informative regularizer has been also used in SVM models considering embedded features selection \citep{watanabe2014disease}. 

In summary, a limited number of contributions have been presented to improve the generalization capability using informative regularizers. In fact, the knowledge added has been primarily focused on spatial information and is insufficient to solve the data shift problem in a time-series domain. Therefore, regularization approaches need to be boosted by means of an injection of high-level knowledge. \color{black}

\section{Methodology}
\label{method}

\subsection{Basic MLP notation}
\label{notation}

Let $\mathcal{N}$ be an artificial neural network with MLP architecture which defines a mapping between two Euclidean spaces $(\mathbb{R}^{n_1}, \mathbb{R}^{n_L}).$ This mapping is obtained by means of a sequence of operators in each layer $i$ and neuron $j$, consisting of an affine transformation

     \begin{align} z_{ij}(x_k) = \sum_{k = 1}^{n_{i-1}}w^i_{jk}x_k - \theta^i_j , \end{align}

\noindent followed by a non-linear function (activation). Multiple alternatives are available to define the activation functions, the most common being the sigmoid activation function, 
     \begin{align} g_{ij}(x) = \sigma \left(z_{ij}(x)\right), \end{align}
    \begin{align}\sigma(z) = \frac{1}{1 + e^{-z}}, \end{align}
    
\noindent and the rectified linear unit (ReLu) activation function, 
    \begin{align}g_{ij}(x) = \max \{0, z_{ij}(x)\}. \end{align}

\color{black}

Note that, in this mathematical model, $g_{ij}(x)$ represents the activation degree in the neuron $j$ of layer $i$.  $n_1, n_2, ..., n_{L}$ are the number of neurons in each layer and $L$ defines the number of layers. $x_k$ is the input value in dimension $k$. The dimensionality of the first layer input is  $\mathbb{D} \in \mathbb{R}^{n\times m}$, where $n$ is the number of features and $m$ the number of objects in the training set (e.g., variable stars). Moreover, each object is defined by the tuple $(d,l)$, where $d \in \mathbb{D}, l \in \mathbb{L}$. 
To train this model, optimization techniques, e.g. mini batch gradient descent, are applied to predict a label $l \in \mathbb{L}$ for each object, minimizing the error in the final mapping with regard to the aforementioned label. The optimization process, typically conducted using a back-propagation variant, searches for a good combination of solutions for each $W^i$ and $\mathbf{\theta}^i $. $W^i \in \mathbb{R}^{n_i \times n_i}$ denotes a weight matrix containing each learnable parameter $w_{jk}^i$ (weights) and $\theta^i$ matrix includes decision variables $\theta^i$ (biases).\color{black}
   
For binary classification, $\mathcal{N}$ has a one-dimensional output layer, in which the activation degree represents the probability of success for this training case.

\subsection{Knowledge injection}
\label{injection}

In the scenario introduced by this paper, training data are shifted and, thus, standard MLP learns from those shifted data. If traditional regularization techniques are applied without human knowledge, the problem remains. Since learning takes place on a non-representative set, only relevant patterns of shifted data are identified. Our paper proposes using an informative regularization approach based on the symbolic representation of well-documented physical knowledge of variable stars to tackle this data shift problem. This representation is based on a characteristic interval of values for an invariant feature $i$ of a variable star $a$, henceforth denoted as $r_i^a$,  that can be defined by:
\begin{equation*}
\text{class } a  \rightarrow \text{  }  x_i \in [l_i, u_i],
\end{equation*}
\noindent where $l_i$ and $u_i$ are the lower and upper characteristic limits for feature $i$. Below we present a set of examples \citep[e.g.,][and references therein]{catelan2014pulsating}:

\begin{itemize}
    \item RR Lyrae $\Rightarrow$ (period $\in [0.2, 1.0]$ days)
    \item RR Lyrae $\Rightarrow$  (amplitude $\in [0.3, 1.2]$ in V-band)
    \item RR Lyrae $\Rightarrow$  (amplitude $\in [0.2, 0.8]$ I-band) 
    \item Classical Cepheid $\Rightarrow$ (period $\in [1, 100]$ days)
    \item type II Cepheid $\Rightarrow$  (period $\in [1, 30]$ days)
\end{itemize}

To inject expert knowledge into the MLP training, we propose to use synthetic inputs called signals. 

\subsubsection{Signal definition}
A signal is defined as a sparse vector (synthetic training data) with non-zero components associated with a set of DRs. Signals contain information about characteristic and short-term invariant features \color{black}(e.g., period and amplitude)\color{black}; the objective herein is to draw samples $s(i)$ for each feature $i$ from a proposed distribution $q$, the parameters of which depend on $r_i^a$.

\color{black} Based on the assumption that $q$ has a uniform distribution, and if we include a perturbation $\epsilon$ on the limits,  we can generate a positive signal (true-class cases) sampling from 
\begin{equation*}
 \underbar{$s$}(i) \sim \mathcal{U}(l_i, l_i+\epsilon) \text{ or}
\end{equation*} 
\begin{equation*}
\bar{s}(i) \sim \mathcal{U}(u_i-\epsilon, u_i);
\end{equation*} 
\noindent a negative signal (false-class) using 
\begin{equation*}
 \underbar{$s$}(i) \sim \mathcal{U}(l_i-\epsilon, l_i) \text{ or }
\end{equation*}
\begin{equation*}
\bar{s}(i) \sim \mathcal{U}(u_i, u_i+\epsilon);
\end{equation*}
\noindent and ambiguous signals by
\begin{equation*}
 \underbar{$s$}(i) \sim \mathcal{U}(l_i-\epsilon, l_i+\epsilon) \text{ or }
\end{equation*}
\begin{equation*}
\bar{s}(i) \sim \mathcal{U}(u_i-\epsilon, u_i+\epsilon).
\end{equation*}
\color{black}

Each signal is represented by a sparse input vector $\mathbf{s} = [0,0,.., s(i),..,0,0]$. The data set that contains those signals is denoted $\mathcal{D}^{\mathbf{s}}$. $s(i)$ contains knowledge only from the DR borders since, based on preliminary experiments, populating the entire DR range does not improve classification performance and requires a considerably larger computational effort. Therefore,  we focus on the DR limits, which are the less represented zones in unidimensional distributions. 

\subsubsection{Bidimensional signals}
\label{2dsignalsdesign}
To generate a bidimensional (2D) signal based on two DRs, \[\text{class } a \rightarrow (x_i \in [l_i, u_i]) \text{ } \wedge \text{ } (x_j \in [l_j, u_j])   \], we propose the following two alternatives for $q$: (i) independent unidimensional (1D) samples can be generated and (ii) a fitting over a  bivariate Gaussian using a subset of data based on DRs, i.e., observation close to the DR borders.  In the latter alternative, to mitigate the data shift problem we apply a shift $\Delta \mu$ over the mean and we assume a full-matrix covariance.  

In      this      case,      the       signal       is        \[ \mathbf{s} = [0,0,.., s(i),.., s(j)..,0,0].\] To select the subset of data, we propose using objects from the class (RR Lyrae); and subsequently, we select a subset of objects from the intersection of the DR bounds. In the case of two rules, the four valid filtering options are presented below.

\begin{itemize}
\item[a)] $(x_i \leq l_i^a +\epsilon_s ) \wedge (x_j \leq l_j^a+\epsilon_s) $
\item[b)] $(x_i \geq u_i^a -\epsilon_s ) \wedge (x_j \geq u_j^a-\epsilon_s)$
\item[c)]  $ (x_i \leq l_i^a +\epsilon_s ) \wedge (x_j \geq u_j^a-\epsilon_s) $
\item[d)] $(x_i \geq u_i^a -\epsilon_s ) \wedge (x_j \leq l_j^a+\epsilon_s) $
\end{itemize}

The $\epsilon_s$ hyperparameter manages the distance from the filtered data to the characteristic DR. Our experiments consider the use of filters a) and b), and the $\epsilon_s$ parameter was set in the range [0.2, 0.4]. Note that, when few objects (i.e., $\sim 100$) are obtained to fit the Gaussian, $\epsilon_s$ must be increased; we recommend an increment of 0.02 each time..\color{black}    

\subsubsection{Discussion about DRs-based signals}
The use of sparse signals helps to inject expert knowledge of characteristic features in under-represented data zones. Moreover, when signals contain few dimensions, human knowledge can be more easily incorporated into proposal distributions.

Traditional approaches for controlling biases or class-imbalance in data based on data augmentation \color{black} may require generating \color{black} large samples of new objects to be effective. In our approach, the training of a subset of weights (a mask) is proposed, whose responsibility is to manage and transmit human knowledge through the neural network. The sparse signals do not need to compete for resources (ANN learning capacity) with training data since the MLP contains a set of weights focused on learning this knowledge. The following section will explain this ad-hoc modeling and the training procedure.

\subsection{Training procedure}
\label{training}

\subsubsection{Bi-objective MLP} 

We optimize a \color{black} binary MLP classifier \color{black} considering a standard bi-objective modeling as follows:
\begin{align}
\centering
  \mathcal{\bar{L}}=& (\mathcal{L}_1, \mathcal{L}_2), \label{f}  \\ 
 \min \mathcal{L}_1 & \label{f1}, \\
 \min \mathcal{L}_2 & \label{f2},
 \end{align}
  \noindent where $\mathcal{L}_1 $ is the binary cross-entropy loss  for training cases, and $\mathcal{L}_2$ is a binary cross-entropy for signals.

In non-shifted data, there is no clear trade-off between the information contained in the data and knowledge from experts; however, in the presence of a data shift problem, we show that structural rules can induce a trade-off and mitigate biases. 

This approach is different to traditional approaches, whose loss functions are weighted as follows:
\begin{align}
\centering
 \min z = \alpha \mathcal{L}_1 + (1- \alpha) \mathcal{L}_2,  \label{fb}
 \end{align}
 
 \noindent where $0 \leq \alpha \leq 1$. In fact, in the weighting sum approach, for each $\alpha$-value in equation (\ref{fb}) we have a solution in the Pareto frontier (non-dominated solutions) for the global bi-objective problem in equations (\ref{f})-(\ref{f2}) \citep{ehrgott2005multicriteria}. In  equation (\ref{fb}), the balance between objective functions is managed by $\alpha$, and sometimes it can be difficult to find the zone where the trade-off in the objectives is generated. \color{black} In our case, class balance is managed by a hyperparameter, $\epsilon_w$, which is a simple threshold of weight values for each loss. Hence, our approach is more intuitive and controllable for assigning priority to each objective depending on the bias level.  \color{black}
 
Consequently, a learning procedure is designed based on two objective functions. The first, $\mathcal{L}_1$, considers a standard classification loss (e.g., cross-entropy). The second, $\mathcal{L}_2$, is focused on fitting the high-level knowledge to obtain a more robust boundary. For $\mathcal{L}_2$, preliminary experiments are conducted using negative, positive and ambiguous signals (probability equal to 0.5). The most significant improvements were obtained by applying a mixture of positive signals (6,000) and negative samples (6,000); therefore, these results are reported.

\subsubsection{Two-step back-propagation} 

A two-step back-propagation algorithm is applied to set the human knowledge in the neural network, in which one step is used for a traditional loss (e.g., cross-entropy) and the other for the regularization loss. Batches for regularization loss contain informative signals and negative examples (objects other than RR Lyrae) from the training set. Adam optimizer was applied for both losses, in which the learning rate for the primary loss was twice that of the auxiliary loss. To avoid the gradient exploding problem, we apply a norm gradient clipping \citep{gradientclipping}.

\subsubsection{Masks} 

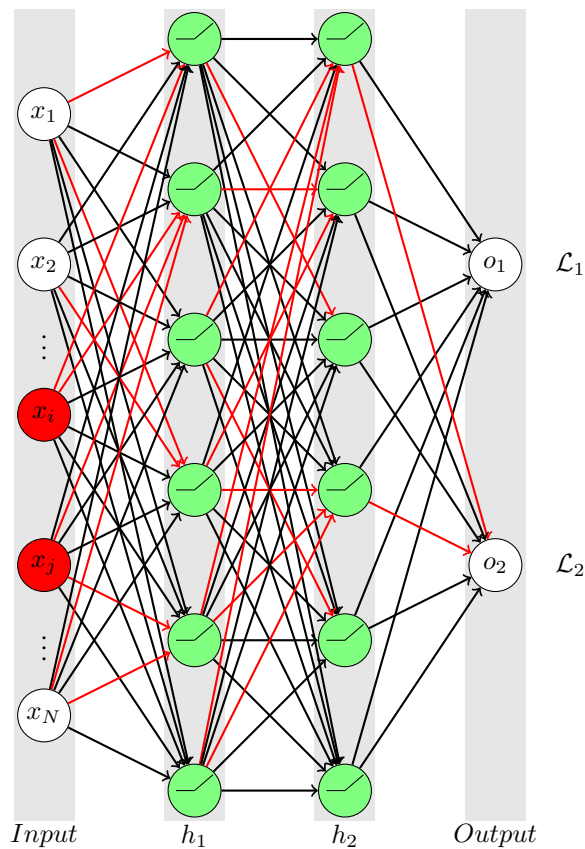
\begin{figure*}
\begin{center}
\tikzset{sigmoid/.style={path picture= {
    \begin{scope}[x=1pt,y=10pt]
      \draw plot[domain=-7:7] (\x,{1/(1 + exp(-\x))-0.5});
    \end{scope}
    }
  }
}

\tikzset{relu/.style={path picture= {
    \begin{scope}[x=7pt,y=6pt]
      \draw plot[domain=-1:0] (\x,0);
      \draw plot[domain=0:1] (\x,\x);
    \end{scope}
    }
  }
}

\tikzstyle{input}=[draw,fill=red!50,circle,minimum size=20pt,inner sep=0pt]
\tikzstyle{hidden}=[draw,fill=green!50,circle,minimum size=20pt,inner sep=0pt]
\tikzstyle{output}=[draw,fill=white,circle,minimum size=20pt,inner sep=0pt]
\tikzstyle{bias}=[draw,dashed,fill=gray!50,circle,minimum size=20pt,inner sep=0pt]
\tikzstyle{layer}=[fill=gray!20]
\tikzstyle{stateTransition}=[->, thick]
\tikzstyle{stateTransition2}=[->, thick, red]
\begin{tikzpicture}[scale=2]
    \fill[layer] (-\width,-3.7) rectangle (\width,1.7);
    \fill[layer] (1+-\width,-3.7) rectangle (1+\width,1.7);
    \fill[layer] (2+-\width,-3.7) rectangle (2+\width,1.7);
    \fill[layer] (3+-\width,-3.7) rectangle (3+\width,1.7);

    \node (l1label) at  (0, -3.8) {$Input$};
    \node (l2label) at  (1, -3.8) {$h_1$};
    \node (l3label) at  (2, -3.8) {$h_2$};
    \node (l4label) at  (3, -3.8) {$Output$};

    \node (r)[input,fill=white]   at (0, 1) {$x_1$};
    \node (g)[input,fill=white] at (0, 0) {$x_2$};
    \node[circle, inner sep=0] (h23) at (0,-0.5) {\vdots};
    \node (b)[input,fill=red]  at (0,-1) {$x_i$};

    \node (l)[input,fill=red]  at (0,-2) {$x_j$};
    \node[circle, inner sep=0] (h23) at (0,-2.5) {\vdots};
    \node (m)[input,fill=white]  at (0,-3) {$x_N$};
    
    \node (h11)[hidden, relu] at (1, 1.5) {};
    \node (h12)[hidden, relu] at (1, 0.5) {};
    \node[hidden, relu] (h13) at (1,-0.5) {};
    \node (h14)[hidden, relu] at (1,-1.5) {};
    \node (h15)[hidden, relu] at (1,-2.5) {};
    \node (h16)[hidden, relu] at (1,-3.5) {};
    
    \node (h21)[hidden, relu] at (2, 1.5) {};
    \node (h22)[hidden, relu] at (2, 0.5) {};
    \node[hidden, relu] (h23) at (2,-0.5) {};
    \node (h24)[hidden, relu] at (2,-1.5) {};
     \node (h25)[hidden, relu] at (2,-2.5) {};
      \node (h26)[hidden, relu] at (2,-3.5) {};

    \node (o1)[output] at (3,0) {$o_1$};
    \node (o2)[output] at (3,-2) {$o_2$};

    \node (l1) at (3.5,0) {$\mathcal{L}_1$};
    \node (l2) at (3.5,-2) {$\mathcal{L}_2$};

    \draw[stateTransition2] (r) -- (h11) node [midway,above=-0.06cm] {};
    \draw[stateTransition] (r) -- (h12) node [midway,above=-0.06cm] {};
    \draw[stateTransition] (r) -- (h13) node [midway,above=-0.06cm] {};
    \draw[stateTransition2] (r) -- (h14) node [midway,above=-0.06cm] {};
    \draw[stateTransition] (r) -- (h15) node [midway,above=-0.06cm] {};
    \draw[stateTransition] (r) -- (h16) node [midway,above=-0.06cm] {};
    
    \draw[stateTransition] (g) -- (h11) node [midway,above=-0.06cm] {};
    \draw[stateTransition] (g) -- (h12) node [midway,above=-0.06cm] {};
    \draw[stateTransition] (g) -- (h13) node [midway,above=-0.06cm] {};
    \draw[stateTransition2] (g) -- (h14) node [midway,above=-0.06cm] {};
    \draw[stateTransition] (g) -- (h15) node [midway,above=-0.06cm] {};
    \draw[stateTransition] (g) -- (h16) node [midway,above=-0.06cm] {};

    \draw[stateTransition2] (b) -- (h11) node [midway,above=-0.06cm] {};
    \draw[stateTransition2] (b) -- (h12) node [midway,above=-0.06cm] {};
    \draw[stateTransition] (b) -- (h13) node [midway,above=-0.06cm] {};
    \draw[stateTransition] (b) -- (h14) node [midway,above=-0.06cm] {};
    \draw[stateTransition] (b) -- (h15) node [midway,above=-0.06cm] {};
    \draw[stateTransition] (b) -- (h16) node [midway,above=-0.06cm] {};
    
    \draw[stateTransition] (l) -- (h11) node [midway,above=-0.06cm] {};
    \draw[stateTransition2] (l) -- (h12) node [midway,above=-0.06cm] {};
    \draw[stateTransition] (l) -- (h13) node [midway,above=-0.06cm] {};
    \draw[stateTransition] (l) -- (h14) node [midway,above=-0.06cm] {};
    \draw[stateTransition2] (l) -- (h15) node [midway,above=-0.06cm] {};
    \draw[stateTransition] (l) -- (h16) node [midway,above=-0.06cm] {};
    
    \draw[stateTransition] (m) -- (h11) node [midway,above=-0.06cm] {};
    \draw[stateTransition2] (m) -- (h12) node [midway,above=-0.06cm] {};
    \draw[stateTransition] (m) -- (h13) node [midway,above=-0.06cm] {};
    \draw[stateTransition] (m) -- (h14) node [midway,above=-0.06cm] {};
    \draw[stateTransition2] (m) -- (h15) node [midway,above=-0.06cm] {};
    \draw[stateTransition] (m) -- (h16) node [midway,above=-0.06cm] {};

    \draw[stateTransition] (h11) -- (h21) node [midway,above=-0.06cm] {};
    \draw[stateTransition] (h11) -- (h22) node [midway,above=-0.06cm] {};
    \draw[stateTransition2] (h11) -- (h23) node [midway,above=-0.06cm] {};
    \draw[stateTransition] (h11) -- (h24) node [midway,above=-0.06cm] {};
    \draw[stateTransition] (h11) -- (h25) node [midway,above=-0.06cm] {};
    \draw[stateTransition] (h11) -- (h26) node [midway,above=-0.06cm] {};
    
    \draw[stateTransition] (h12) -- (h21) node [midway,above=-0.06cm] {};
    \draw[stateTransition2] (h12) -- (h22) node [midway,above=-0.06cm] {};
    \draw[stateTransition] (h12) -- (h23) node [midway,above=-0.06cm] {};
    \draw[stateTransition] (h12) -- (h24) node [midway,above=-0.06cm] {};
    \draw[stateTransition] (h12) -- (h25) node [midway,above=-0.06cm] {};
    \draw[stateTransition] (h12) -- (h26) node [midway,above=-0.06cm] {};
    
    \draw[stateTransition2] (h13) -- (h21) node [midway,above=-0.06cm] {};
    \draw[stateTransition] (h13) -- (h22) node [midway,above=-0.06cm] {};
    \draw[stateTransition] (h13) -- (h23) node [midway,above=-0.06cm] {};
    \draw[stateTransition] (h13) -- (h24) node [midway,above=-0.06cm] {};
    \draw[stateTransition2] (h13) -- (h25) node [midway,above=-0.06cm] {};
    \draw[stateTransition] (h13) -- (h26) node [midway,above=-0.06cm] {};
    
    \draw[stateTransition2] (h15) -- (h21) node [midway,above=-0.06cm] {};
    \draw[stateTransition] (h15) -- (h22) node [midway,above=-0.06cm] {};
    \draw[stateTransition] (h15) -- (h23) node [midway,above=-0.06cm] {};
    \draw[stateTransition2] (h15) -- (h24) node [midway,above=-0.06cm] {};
    \draw[stateTransition] (h15) -- (h25) node [midway,above=-0.06cm] {};
    \draw[stateTransition] (h15) -- (h26) node [midway,above=-0.06cm] {};

    \draw[stateTransition] (h14) -- (h21) node [midway,above=-0.06cm] {};
    \draw[stateTransition2] (h14) -- (h22) node [midway,above=-0.06cm] {};
    \draw[stateTransition] (h14) -- (h23) node [midway,above=-0.06cm] {};
    \draw[stateTransition2] (h14) -- (h24) node [midway,above=-0.06cm] {};
    \draw[stateTransition] (h14) -- (h25) node [midway,above=-0.06cm] {};
    \draw[stateTransition] (h14) -- (h26) node [midway,above=-0.06cm] {};
    
    \draw[stateTransition2] (h16) -- (h21) node [midway,above=-0.06cm] {};
    \draw[stateTransition] (h16) -- (h22) node [midway,above=-0.06cm] {};
    \draw[stateTransition] (h16) -- (h23) node [midway,above=-0.06cm] {};
    \draw[stateTransition2] (h16) -- (h24) node [midway,above=-0.06cm] {};
    \draw[stateTransition] (h16) -- (h25) node [midway,above=-0.06cm] {};
    \draw[stateTransition] (h16) -- (h26) node [midway,above=-0.06cm] {};
    
    \draw[stateTransition] (h21) -- (o1) node [midway,above=-0.06cm] {};
    \draw[stateTransition] (h22) -- (o1) node [midway,above=-0.10cm] {};
    \draw[stateTransition] (h23) -- (o1) node [midway,above=-0.06cm] {};
    \draw[stateTransition] (h24) -- (o1) node [midway,above=-0.06cm] {};
    \draw[stateTransition] (h25) -- (o1) node [midway,above=-0.06cm] {};
    \draw[stateTransition] (h26) -- (o1) node [midway,above=-0.06cm] {};

    \draw[stateTransition2] (h21) -- (o2) node [midway,above=-0.06cm] {};
    \draw[stateTransition] (h22) -- (o2) node [midway,above=-0.10cm] {};
    \draw[stateTransition] (h23) -- (o2) node [midway,above=-0.06cm] {};
    \draw[stateTransition2] (h24) -- (o2) node [midway,above=-0.06cm] {};
    \draw[stateTransition] (h25) -- (o2) node [midway,above=-0.06cm] {};
    \draw[stateTransition] (h26) -- (o2) node [midway,above=-0.06cm] {};

\end{tikzpicture}
\end{center}
\caption{ Bi-objective MLP with masks for each objective. Each gray rectangle contains a layer. In the input layer, red circles represent the non-zero values in each signal. Green circles in hidden layers $h_1$ and $h_2$ represent neurons. Arrows show the flow of information through the network. The arrow color indicates the loss function from where the weight update comes from, where  $\mathcal{L}_1$ represents the primary objective which is related to the classification task, and $\mathcal{L}_2$ is the regularization loss, which is focused on adding human knowledge into the training. Red weights are updated using the propagated error from $\mathcal{L}_2$ and black weights are updated using the back-propagation from $\mathcal{L}_1$. \color{black}}
\label{netdiag}
\end{figure*}

An iterative back-propagation is implemented with masks focused on specific losses, i.e., iterations 1, 3, 5, . . . , $2k+1$ are focused on optimizing $\mathcal{L}_1$ and iterations 2, 4, 6, . . . , $2k$ are focused on $\mathcal{L}_2$. Each loss has a binary mask for the back-propagation step. The masks can be understood as two independent systems from a modeling perspective. A primary system, which is based on training data, aims to memorize the class for each data belonging to the training set. A secondary system is responsible for incorporating structural knowledge into the learning process. An initial training phase is performed to assign the weights to each mask (\color{black} first 50 epochs of training\color{black}). In this phase, the weights can change of assigned mask (assigned loss function) in each epoch. The weights smaller than a threshold, $\epsilon_w$, are assigned to $\mathcal{L}_2$, and the remaining weights are optimized considering $\mathcal{L}_1$. Smaller weights ($\leq |\epsilon_w| $) are assigned to signal-based loss ($\mathcal{L}_2$); in other words, weights with low relevance for activating or inhibiting neurons are optimized for the auxiliary loss of regularization. Following this initialization phase, the weights are fixed to a loss. To provide a more stable training procedure, bias parameters ($\theta_j$) and batch normalization learnable parameters are assigned to the primary system. \color{black} Based on the convergence behavior and testing performance, hyperparameter $\epsilon_w$ was set to $0.025$ and hyperparameter $\epsilon_s$ was set to $0.2$. Figure \ref{netdiag}  provides a visualization of the training scheme, in which the features that receive human-based information from signals are colored red. Green circles represent activation functions. Connections (weights) between nodes that are optimized via the application of the back-propagation from $\mathcal{L}_2$ are colored red The remaining weights are updated by using back-propagation from $\mathcal{L}_1$ and are colored black. Figure \ref{convergence} shows a sample of the convergence behavior of our training procedure. 

\subsubsection{Other hyperparameters} 

In preliminary experiments, batch sizes of 32, 64, 128, 256 and 512 are used. The best results were obtained with a batch size of 256; hence only these are reported. ReLU activation functions are always used throughout. An early stopping strategy using a patience parameter equal to ten epochs is implemented, i.e., the validation loss checking is carried out every ten epochs. The maximum number of epochs is 1,500. Due to the imbalance problem in the training set, a downsampling approach is applied. \color{black} Our downsampling first creates a balanced set; it is to say, we sample from the majority class $n^\prime$ objects, where $n^\prime$ is the number of samples in the minority class (true class); once we have a balanced data set, we apply a new sample of $n$ objects (e.g., 5,000, 10,000 or 50,0000) for the final training set of each experiment. 

To select hyperparameters in preliminary experiments, we use a grid search method. Each of those hyperparameters is relevant and has an impact on the final results, but focusing on generalization, it is essential to highlight the early stopping criteria. The patience hyperparameter allows managing the overfitting; when a higher patience value is used, the overfitting is increased, and obviously, the performance in the testing set is decreased. Another relevant pattern was observed in the batch size; the impact of the size batch on optimization results has been studied. The evidence shows that a small batch size generates flat minima (better generalization), and on the other hand, bigger batches tend to lead to sharp minima \citep{keskar2016large}. However, this evidence is based on a representative training set, so each batch is a good sample; in a case considering data shift, that is not true, and because of that, small batches can provide very wrong directions for the optimizer. In our preliminary experiments, small batches provided models with worse generalization, and sets of 256 and 512 objects produced similar results; hence, to prioritize a flat minimum \citep{keskar2016large}, we chose 256 objects in each batch.
\color{black}

\color{black}

\section{Data}
\label{data}

\subsection{OGLE-III Catalog of Variable Stars}
\label{ogle}

The OGLE-III Catalog of Variable Stars was used to assess the methodology applied in this paper. As its name implies, the catalog corresponds to the third phase of the project \citep{udalski2008optical}. The main goal of OGLE is to identify microlensing events and transiting exoplanets in four fields: the Galactic bulge, the Large and Small Magellanic Clouds, and the constellation of Carina. 

\subsection{Processing of light curves}
\label{designmatrix}

The corresponding features of each light curve were extracted using the Feature Analysis for Time Series (FATS) library \citep{nun2015fats}. \color{black} This study used OGLE light curves (I band) with at least 25 observations. A sample of 300 observations was used for each light curve with more than 300 observations. Objects with outliers (i.e., values beyond the mean $\mp$ three standard deviation) in features were discarded, reducing the true-class by less than 2\%. \color{black} Thus, a matrix of 401,186$\times$60 was obtained, where 60 stands for the number of separate features included in the analysis \color{black}(see these features in Table \ref{features}) \color{black}. These features will be used to experiment with the aforementioned training procedure for MLPs.

\subsection{Shifted data}
\label{shifteddata}
To validate the approach used in this study for mitigating the data shift problem, we apply the method for shifting data proposed by \citet{perez2021informative}. \color{black} In this method, first, a classifier is trained using the complete set of labeled objects; after that, a soft prediction is conducted; and finally, a biased selection procedure assigns objects to two data sets (A and B). This bias considers that more easily separable objects are assigned to the A set more frequently, and consequently, more challenging classifications are assigned more frequently to the B set. Then, two shifted data sets are obtained; we consider A for the training set and B for the testing set, emulating the underlying biases in variable stars.  \color{black}

Following a description of the aforementioned pre-processing, this section moves on to discuss the distribution of variable star types in each set, which is presented in Table \ref{tab:my-table}. \color{black} As we can see in Table \ref{tab:my-table}, the obtained sub-class distribution is shifted when comparing training and testing sets; the presence of RRab in training ($\sim$77\%) is higher than in the testing set ($\sim$60\%), on the contrary, the percentage of RRc and RRd in testing ($\sim$30\% and $\sim$6\%) is higher than in the training set ($\sim$18\% and $\sim$2\%). Discrepancies can also be observed (see Table \ref{tab:my-table2}) when we study the proportion of true-class objects in each field.

 Two levels of information are visualized for both sets. On the first level, Figure \ref{samplelightcurves} shows a sample of light curves belonging to the training and testing sets. On the second level, Figure \ref{densityplot} contrasts training and testing sets over the space of relevant features. The relevance of these features was estimated using the mean decrease in impurity from an RF classifier. \color{black} According to this procedure, the most important features are: period  (PeriodLS),  amplitude, median absolute deviation (MedianAbsDev), interquartile range (Q31), mean, standard deviation (Std), autocorrelation length given a threshold (Autocor\_length), and amplitude of the first harmonic in the first frequency (Freq1\_harmonics).

\begin{figure*}
\centering
\includegraphics[scale=0.55]{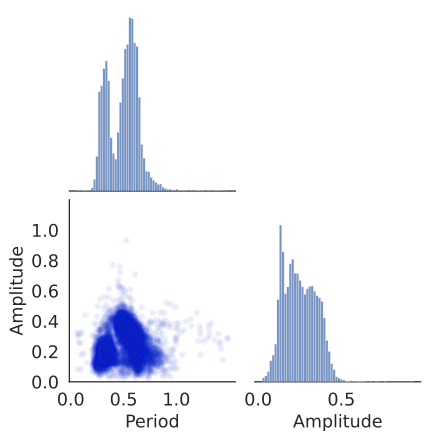}\hspace{0mm}
\includegraphics[scale=0.55]{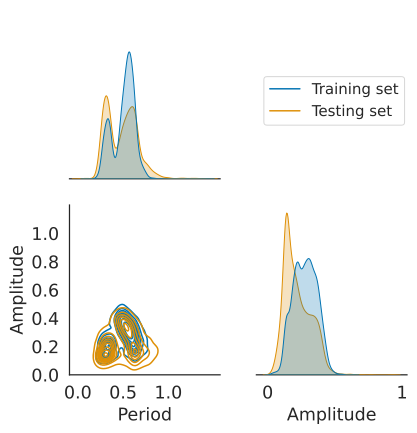}\hspace{0mm}
\caption{\color{black} Period and amplitude density distributions for training and testing sets. The plot on the left shows the univariate and bivariate density of these features, merging testing and training sets. The plot on the right shows the same distributions but splitting testing and training sets.   We apply the kernel density estimation method to provide these visualizations using a sample of 5,000 objects in each data set. \color{black}}
\label{densityplotb}
\end{figure*}

\begin{figure*}
\begin{minipage}{165mm}
\begin{subfigure}{.5\textwidth}
  \centering
  \includegraphics[width=1\linewidth]{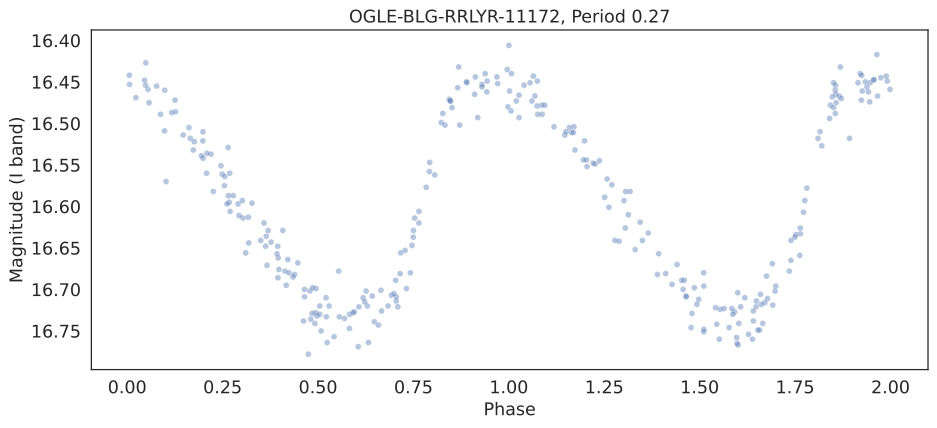}
  \label{fig:sfig1a}
\end{subfigure}%
\begin{subfigure}{.5\textwidth}
  \centering
  \includegraphics[width=1\linewidth]{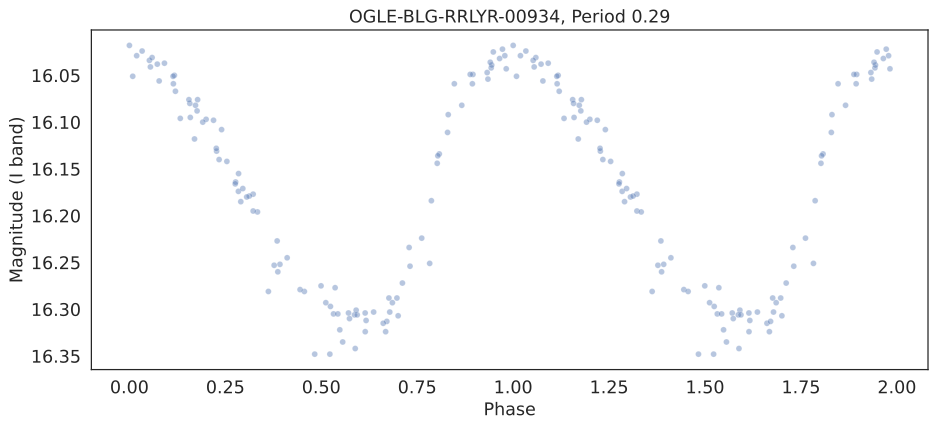}
  \label{fig:sfig2b}
\end{subfigure}
\begin{subfigure}{.5\textwidth}
  \centering
  \includegraphics[width=1\linewidth]{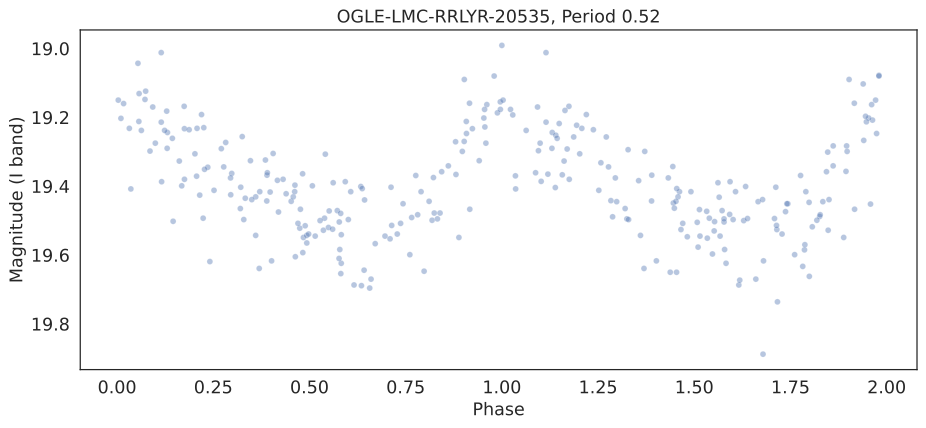}
  \label{fig:sfigb6}
\end{subfigure}%
\begin{subfigure}{.5\textwidth}
  \centering
  \includegraphics[width=1\linewidth]{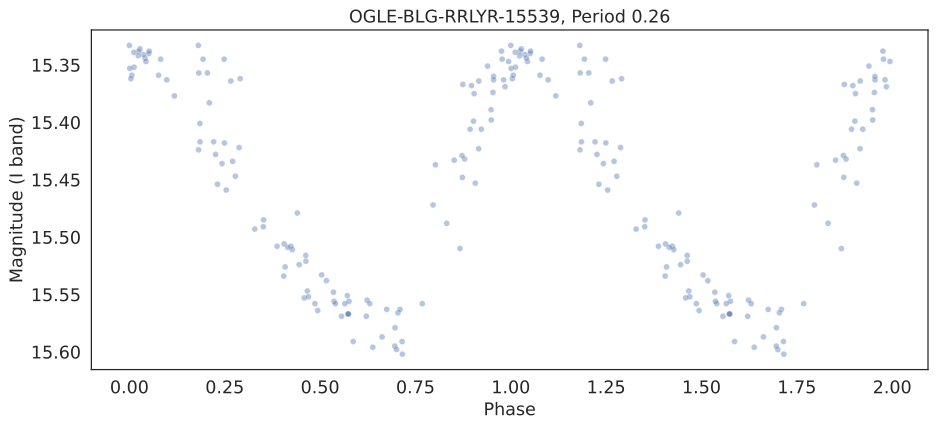}
  \label{fig:sfig3b}
\end{subfigure}
\begin{subfigure}{.5\textwidth}
  \centering
  \includegraphics[width=1\linewidth]{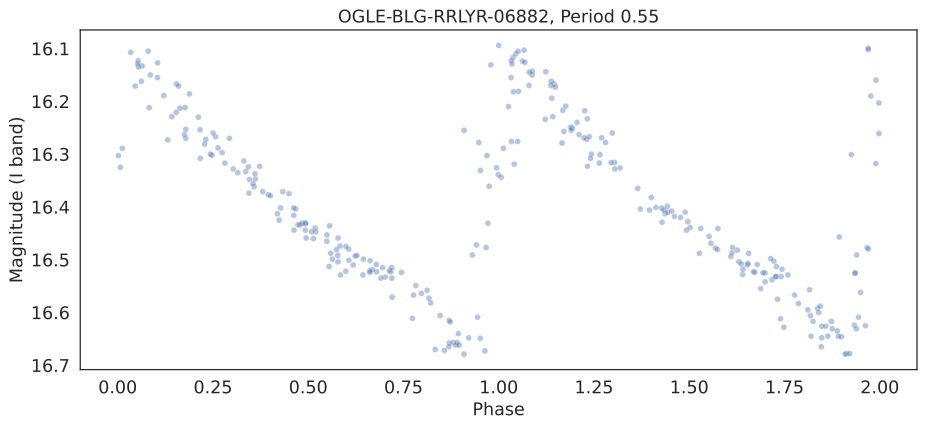}
  \label{fig:sfig4}
\end{subfigure}%
\begin{subfigure}{.5\textwidth}
  \centering
  \includegraphics[width=1\linewidth]{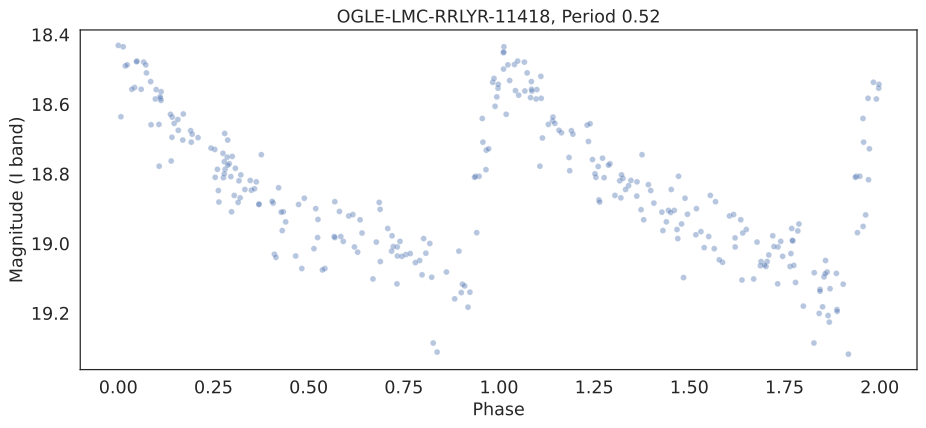}
  \label{fig:sfig5bb2}
\end{subfigure}
\begin{subfigure}{.5\textwidth}
  \centering
  \includegraphics[width=1\linewidth]{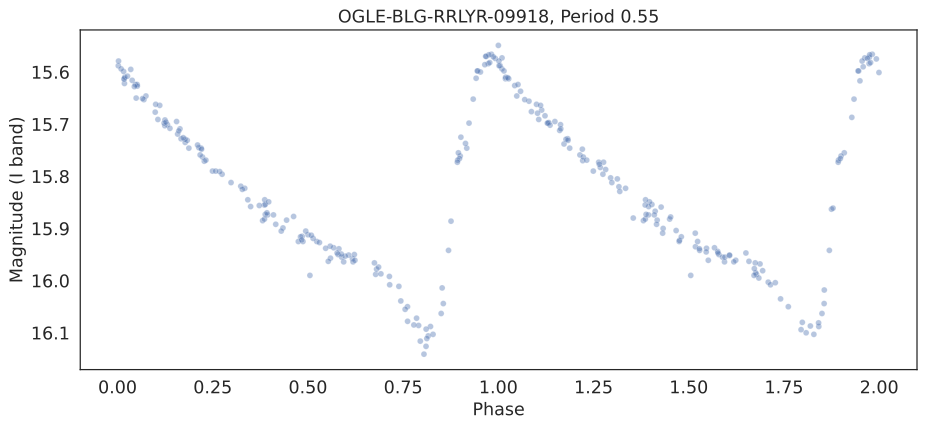}
  \label{fig:sfig4b}
\end{subfigure}%
\begin{subfigure}{.5\textwidth}
  \centering
  \includegraphics[width=1\linewidth]{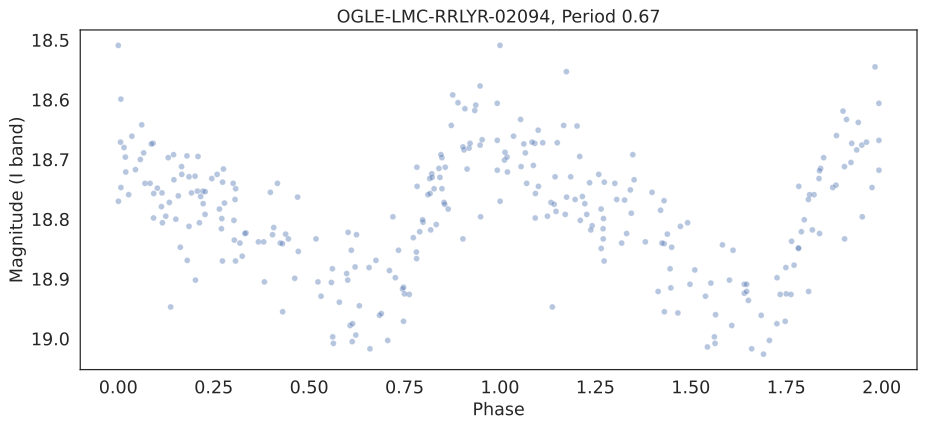}
  \label{fig:sfig5b}
\end{subfigure}
\begin{subfigure}{.5\textwidth}
  \centering
  \includegraphics[width=1\linewidth]{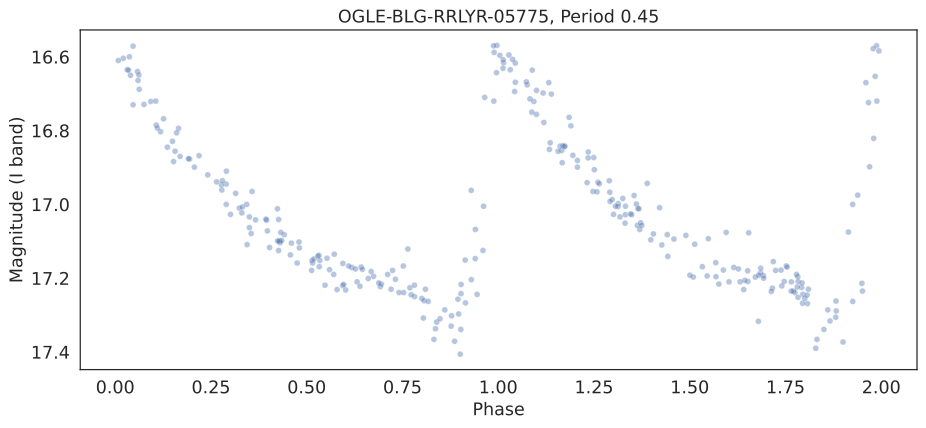}
  \label{fig:sfig4bb}
\end{subfigure}%
\begin{subfigure}{.5\textwidth}
  \centering
  \includegraphics[width=1\linewidth]{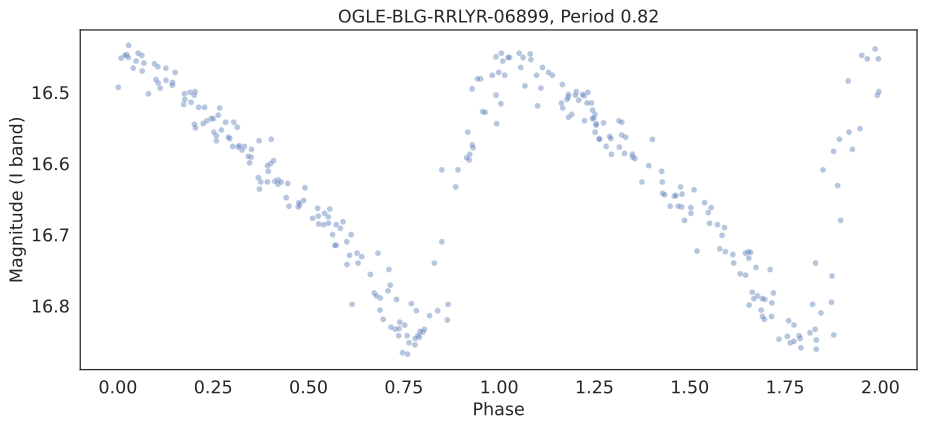}
  \label{fig:sfig5bb}
\end{subfigure}
\caption{\color{black} Sample of phased light curves of RR Lyrae variable stars. The stars to the left relate to the training set and the stars to the right relate to the testing set. At the top of each panel, the star's OGLE ID is provided, along with its period (in days). The OGLE ID includes the field (see Table \ref{tab:my-table2}) and star number as well. \color{black}
} 
\label{samplelightcurves}
\end{minipage}
\end{figure*}

\begin{figure*}
\centering
\includegraphics[width=17cm, height=17cm]{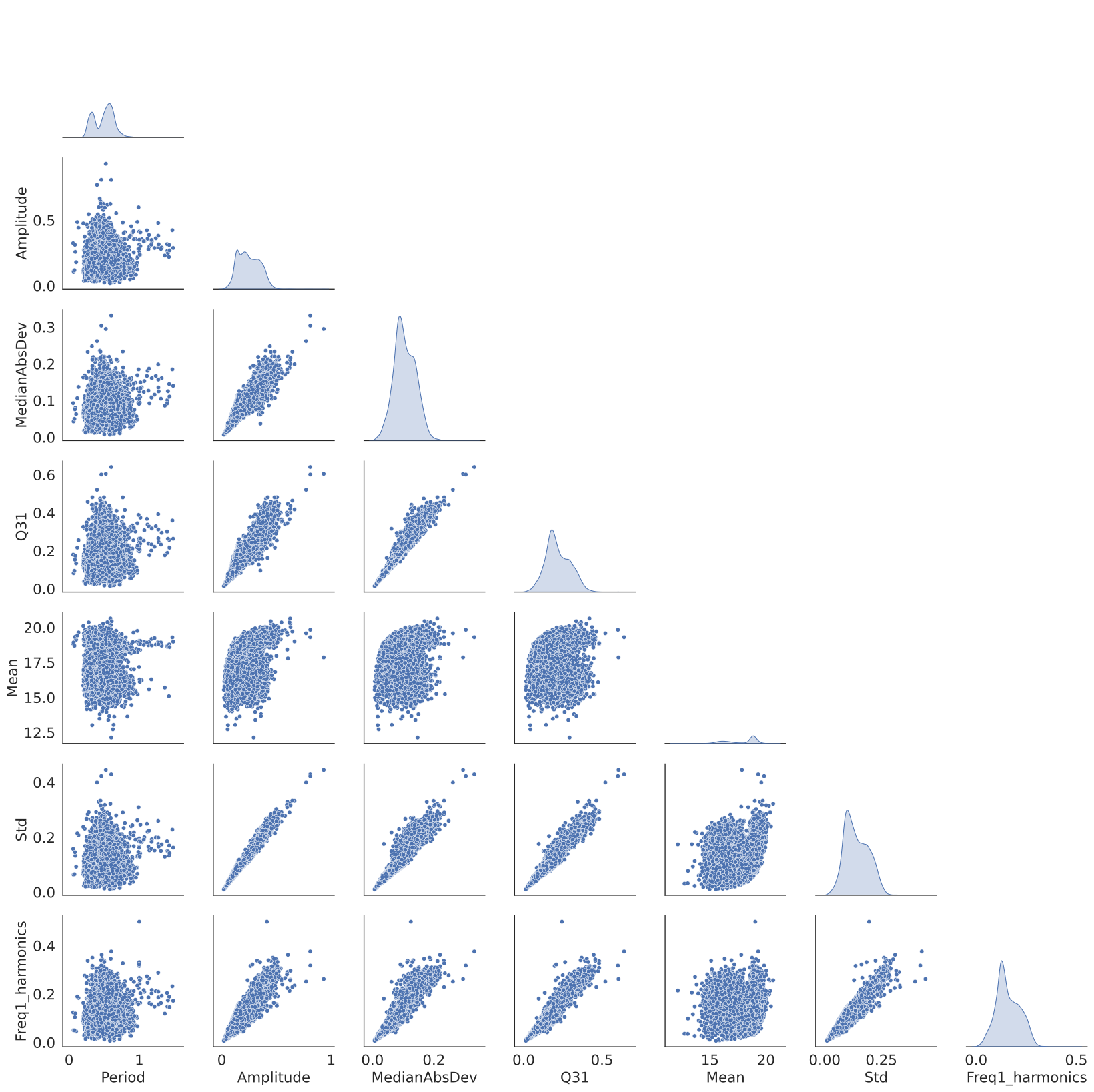}\hspace{0mm}
\caption{\color{black} The main-diagonal subplots show a univariate density fit of each feature;  whereas the plots below the main-diagonal show the bivariate density fit contour for each set. \color{black}}
\label{densityplot}
\end{figure*}

\begin{figure*}
\centering
\includegraphics[width=17cm, height=17cm]{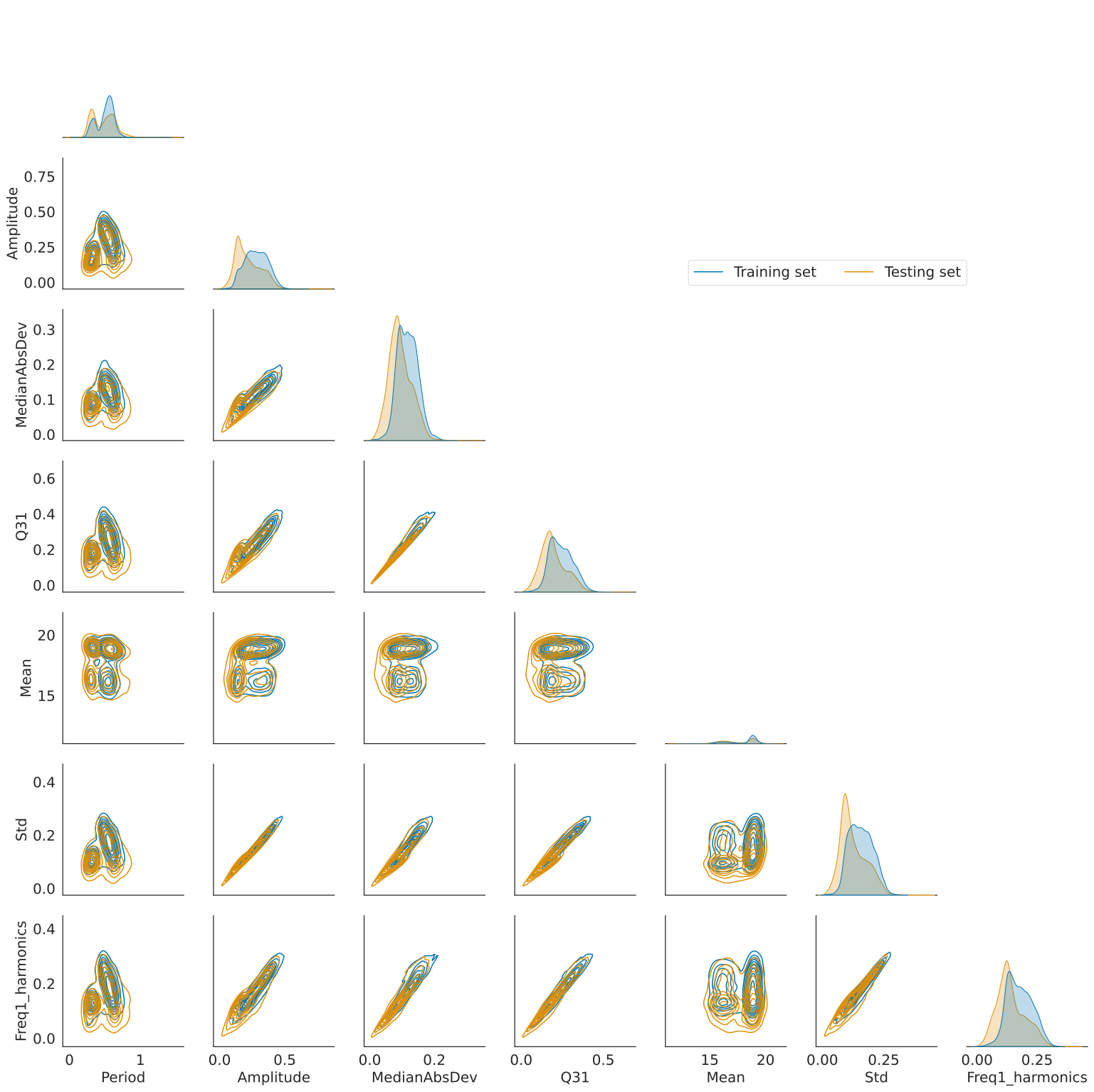}\hspace{0mm}
\caption{\color{black} As in Figure~\ref{densityplot} but splitting training and testing sets (shown in black and orange, respectively). \color{black}}
\label{densityplot2}
\end{figure*}

When observing Figure \ref{samplelightcurves}, detecting the shift between light curves belonging to training and testing sets is not straightforward since there are multiple patterns and points in each light curve. \color{black} However, when the feature space is observed in Figures \ref{densityplotb} and \ref{densityplot2}, the data shift is visualized; for example, we can observe the shift between training and testing means. \color{black}

\color{black} Figure \ref{densityplotb} provides visualizations of the period and amplitude density distribution for RR Lyrae stars. Figure \ref{densityplotb}(left) provides density distributions for the complete labeled set (i.e., merging testing and training sets) and Figure \ref{densityplotb}(right) shows the same distributions but splitting both sets. The contrast between these plots gives clear evidence of the induced data shift. In both figures, univariate distributions for period and amplitude are considered in main-diagonal pannels. The lower left panel includes a merge of objects from both sets and provides the joint density distribution, also known as \textit{Bailey diagram} \citet{catelan2014pulsating}. When looking Figure \ref{densityplotb}(right), the shift is identified in the univariate case; for example, the same two populations are observed in univariate period density, but the balance between modes is shifted. In the case of univariate amplitude density, the mode of the first population and the relative representation of each population are also shifted.  From Bailey diagrams, we observe the classic bimodal distribution where RR Lyrae stars with the longest periods at any given amplitude are RRab. RR Lyrae stars with short periods and small amplitudes are RRc. From these visualizations and Table \ref{tab:my-table}, we can observe that RRab stars have a higher concentration in the training set (population with a longer period); on the contrary, RRc stars have a higher concentration in the testing set.   \color{black}

Figures \ref{densityplot} and \ref{densityplot2} provide the same visualizations for the seven most relevant features previously introduced. The shift is identified in each univariate density since the kurtosis is constantly shifted and, in certain cases, e.g. amplitude, the means also differ. In bivariate plots below the main diagonal, most of the density mass is observed in at least two populations. \color{black}

\begin{table*}
\color{black}
\centering
\caption{Training and testing class distribution from OGLE-III labeled set. }
\begin{tabular}{lll}
\textbf{Class}                                       & \textbf{Testing objects} & \textbf{Training objects} \\ \hline
\multicolumn{1}{l|}{\textbf{Total}}                        &  27,293          &    373,893        \\
\multicolumn{1}{l|}{\textbf{}}                        &            &            \\
\multicolumn{1}{l|}{\textbf{True class (RR Lyrae)}}    & 13,694            & 28,243 \\
\multicolumn{1}{l|}{RRab}                        &      8,156 (59.56\%)    &     21,637   (76.61\%)        \\
\multicolumn{1}{l|}{RRc}                        &      4,218 (30.80\%)    &        5,078 (17.98\%)      \\
\multicolumn{1}{l|}{RRd}                        &     493   (3.60\%)     &    974 (3.45\%)        \\
\multicolumn{1}{l|}{RRe}                        &    827  (6.04\%)     &       554  (1.96\%)      \\
\multicolumn{1}{l|}{\textbf{}}                        &            &            \\

\multicolumn{1}{l|}{\textbf{False class  (non-RR Lyrae) }}                        &      13,599      &    345,650        \\
\multicolumn{1}{l|}{Eclipsing Binary}                & 6,383    (46.94\%)         & 30,859    (8.93\%)         \\
\multicolumn{1}{l|}{Cepheids}                        & 3,844    (28.27\%)         & 4,061      (1.17\%)        \\
\multicolumn{1}{l|}{Long-Period Variables}           & 1,956  (14.38\%)            & 308,997  (89.40\%)          \\
\multicolumn{1}{l|}{Delta Scuti}                     & 1,125    (8.27\%)          & 1,255   (0.36\%)            \\
\multicolumn{1}{l|}{Type II Cepheid}                 & 215    (1.51\%)          & 348     (0.1\%)          \\
\multicolumn{1}{l|}{Anomalous Cepheid}               & 45      (0.33\%)         & 35         (0.01\%)       \\
\multicolumn{1}{l|}{Double Periodic Variable}        & 26       (0.19\%)           & 100       (0.03\%)         \\
\multicolumn{1}{l|}{Dwarf Nova}                      & 5         (0.03\%)           & 2     ($< $0.01 \%)            \\
\multicolumn{1}{l|}{$\alpha^2$ Canum Venaticorum}    & 0                & 5              ($< $0.01 \%)    \\
\multicolumn{1}{l|}{R CrB Variable}                  & 0                & 7          ($< $0.01 \%)  \\    \hline
\end{tabular}
\label{tab:my-table}
\end{table*}
\color{black}

\begin{table*}
\caption{\color{black} Distribution of field for true-class objects (RR Lyrae) in training and testing sets. }
\color{black}
\centering
\begin{tabular}{l|r|r}
\hline
Field & Training objects & Testing objects \\
\hline
   Galactic Bulge (BLG)    &   47.57\%        &   61.57\%       \\
  Large Magellanic Cloud (LMC)    &    47.33\%      &     32.31\%      \\
     Small Magellanic Cloud (SMC)  &    4.91\%      &    6.11\%      \\
  Galactic Disk (GD)    &       0.19\%  &       0.02\%    \\
\hline
\end{tabular}
\label{tab:my-table2}
\end{table*}
\color{black} 

\begin{table*}
\centering
\caption{Summary of model's performance decay from training set to testing set (see Table \ref{tab:my-table}). Each experiment represents the mean value on the testing set from a baseline model trained using a random sampling approach within the \color{black} full \color{black} training set. To avoid overfitting, an early stopping strategy is considered using a validation set from the training set.}
\begin{tabular}{lrrrrrr}
\toprule
&        \multicolumn{2}{c}{Accuracy} & \multicolumn{2}{c}{F1 score} & \multicolumn{2}{c}{AUC} \\  
&        training & testing & training & testing & training & test \\     
\midrule
5,000     &     99.78 &    77.42 &      99.87 &   77.42 &       1.00 &     0.84 \\
10,000   &     99.81 &    78.87 &      99.88 &   79.15 &       1.00 &     0.85 \\
50,000   &     99.87 &    77.49 &      99.91 &   82.98 &       1.00 &     0.85 \\
\bottomrule
\end{tabular}
\label{decay}
\end{table*} 

Consequently, these induced biases generate a performance decay on the trained models. Table \ref{decay} shows this detriment, considering the following three metrics: 

\begin{itemize}
    \item \textbf{Accuracy:} This evaluates the prediction quality on the testing set based on the ratio of correct predictions over the total number of observations. We consider a threshold equal to 0.5 for defining the true class.
    \item \textbf{F1-score:} To include a balance between recall (i.e., true positive rate) and precision (i.e., positive predictive rate), we consider the F1-score, which is the harmonic mean between these, in addition, this is more robust than accuracy for imbalanced classes. 
    \item \textbf{AUC:} Unlike accuracy and F1 metrics, AUC (area under the receiver-operating characteristic curve) is based on the model performance for each possible classification threshold \citep{davis2006relationship}. The receiver-operating characteristic curve provides the true positive rate with respect to the false positive rate, being an AUC equal to 1.0 for a perfect model score and 0.5 in the case of a random model. AUC can be understood as the probability that the assessed model classifies a random true-class object more probably than a random false-class object.   
\end{itemize}

As we can see, the model's performance in the three metrics is perfect on training sets, getting the maximum score (or very close to the maximum). However, these metrics decay when the model is tested beyond the training set, even applying traditional strategies to mitigate overfitting. The baseline model used in these experiments is an MLP with the hyperparameters (and architecture) presented in Section \ref{training} but without our informative regularization approach.        

\section{Implementation}
\label{implementation}

To implement the training procedure and the knowledge injection approach, \textsc{Python 3.7} was used. The MLP was implemented using \textsc{Pytorch 1.0.1} \citep{NEURIPS2019_9015}. Additional libraries used include \textsc{Scikit-Learn} \citep{pedregosa2011scikit}, \textsc{Pandas}  \citep{mckinney2011pandas} and \textsc{Seaborn}    \citep{waskom2021seaborn}.  Finally, the source code is available at \href{https://github.com/frperezgalarce/vs\_mlp}{https://github.com/frperezgalarce/vs\_mlp}. 

\section{Experiments and results}
\label{results}

\subsection{Regularization using unidimensional knowledge injection}
\label{1d}

To assess the impact of our proposal, when 1D signals are injected during the MLP training, we compare the following four approaches: (i) an MLP without regularization, the baseline model; (ii) the baseline model using dropout as regularization; and two (iii-iv) informative regularization approaches using our training procedure. Our proposal includes signals based on expert knowledge from period and amplitude features \color{black} (I band), \color{black} and we refer to them as 1D-period and 1D-amplitude, respectively. \color{black} 

\begin{figure*}
\centering
\includegraphics[scale=0.4]{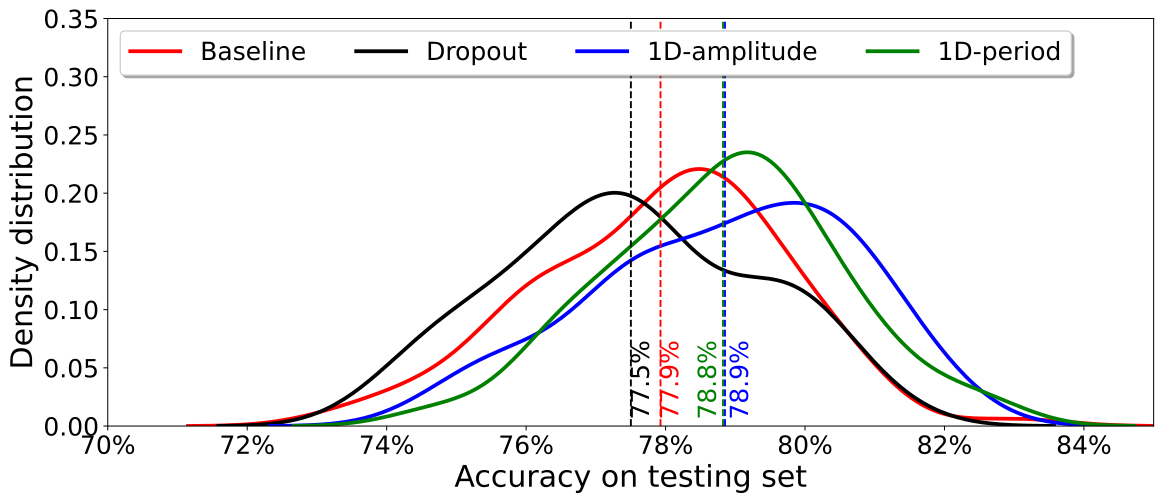}\hspace{0mm}
\caption{ Density distribution for the accuracy in the testing set; vertical lines show mean accuracy for each training execution. Thirty experiments for each alternative were conducted, considering 5,000, 10,000 and 50,000 training cases. \color{black} Each experiment considers a model trained using a random sampling approach within the training set, we apply the same approach for the following results. \color{black}  The x-axis is truncated to have a range from 70\% to 85\%.}
\label{figure3}
\end{figure*}

Figure \ref{figure3} shows the distribution of the test-set accuracy for 30 experiments. When the informative loss is activated, we note a bigger concentration toward high accuracy values. Moreover, the mean accuracy is also improved (from 77.9\% to 78.9\%) when the informative loss is used.  \color{black} We apply t-Student tests for independent samples to check the statistical significance of these results, using a Shapiro-Wilk test to validate normality. In both cases, the null hypothesis was rejected ($\alpha =0.05$), thus implying that the differences in test-set accuracy are statistically significant. Mann-Whitney-Wilcoxon test and Kolmogorov-Smirnov test also confirmed this result.\color{black}  

We apply t-Student tests for independent samples to check the statistical significance of these results, using a Shapiro-Wilk test to validate normality. In both cases, the null hypothesis was rejected ($\alpha =0.05$), thus implying that the differences in test-set accuracy are statistically significant. A Mann-Whitney-Wilcoxon test also confirmed this result.

\begin{figure*}
\centering
\subfloat[]{\includegraphics[scale=0.4]{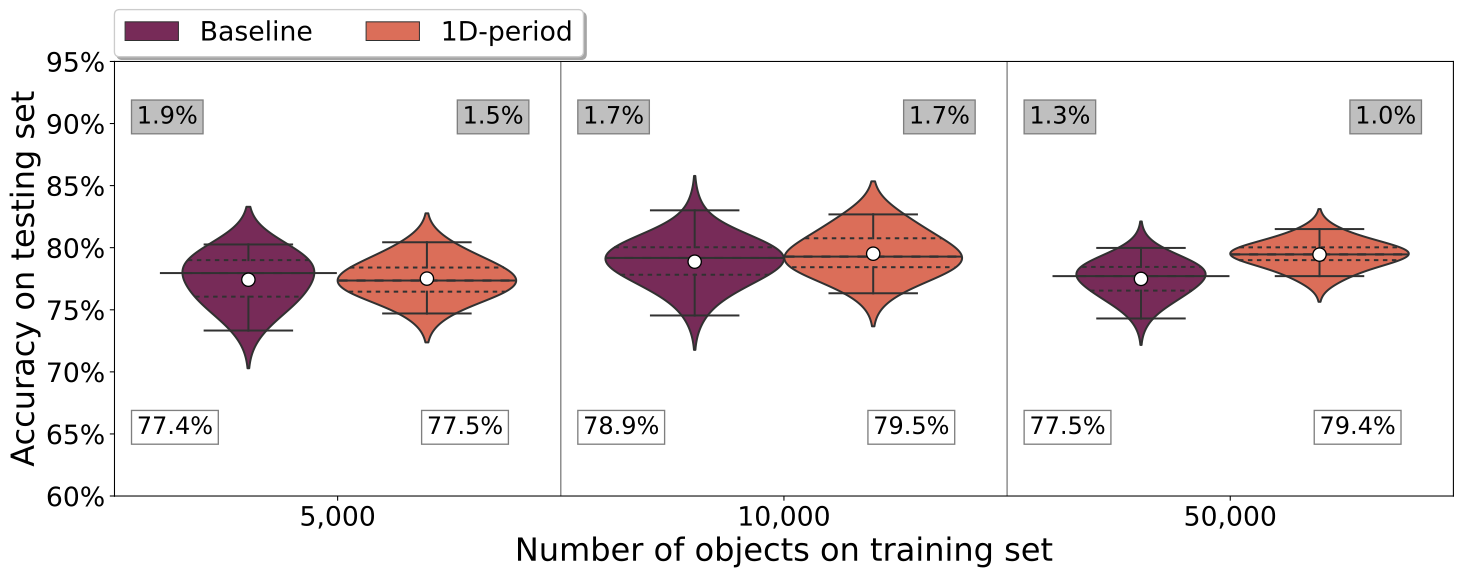}}\hspace{0mm}
\subfloat[]{\includegraphics[scale=0.4]{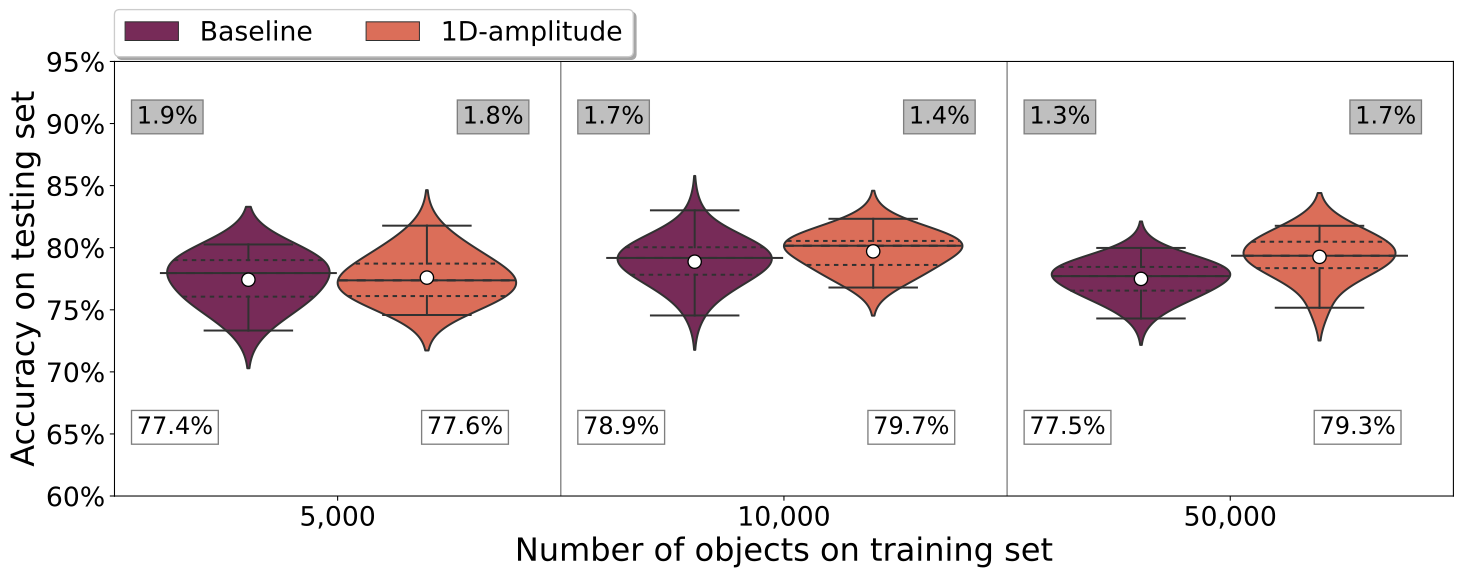}}\hspace{0mm}
\caption{Impact of regularization using 1D signals on the accuracy distribution utilizing 256 objects per batch. White points represent the mean, black lines show the median, and dotted lines indicate the first and third quartiles. Gray boxes show the standard deviation value, whereas white boxes contain the mean value. Thirty experiments for each plot were undertaken. (a) provides results for 1D-period signals and (b) for 1D-amplitude signals. The y-axis is truncated to have a range from 60\% to 95\%. \color{black}}
\label{periodamplitud}
\end{figure*}

Figure \ref{periodamplitud} shows the impact of our regularization method injecting 1D signals on the accuracy. White points represent the mean, black lines show the median, and dotted lines indicate the first and third quartiles. The gray boxes show the standard deviation, and the white boxes contain the mean. Thirty experiments for each plot were carried out. 

Figure \ref{periodamplitud}(a) provides results for 1D-period signals and Figure \ref{periodamplitud}(b) for 1D-amplitude signals. When looking at Figure  \ref{periodamplitud}(a), we can observe that when the median (white boxes) is considered, our 1D-period signal approach always outperforms the baseline model. Similarly, Figure \ref{periodamplitud}(b) presents the results for 1D-amplitude signals, where an improvement is also generated. The biggest improvement (1.8\%) was obtained using 1D-period considering 50,000 training cases, being statistically significant according to the t-Student’s test.

\subsection{Regularization using bidimensional knowledge injection}
\label{2d}

\begin{figure*}
\centering
\subfloat[]{\includegraphics[scale=0.4]{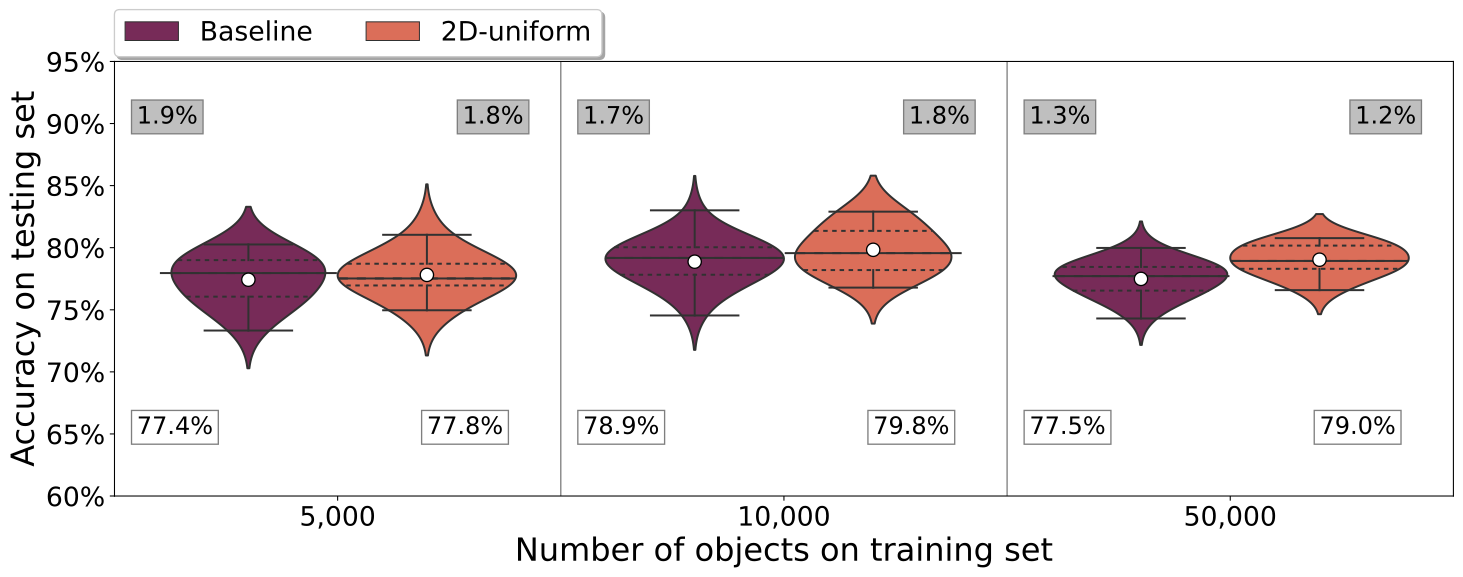}}\hspace{0mm}
\subfloat[]{\includegraphics[scale=0.32]{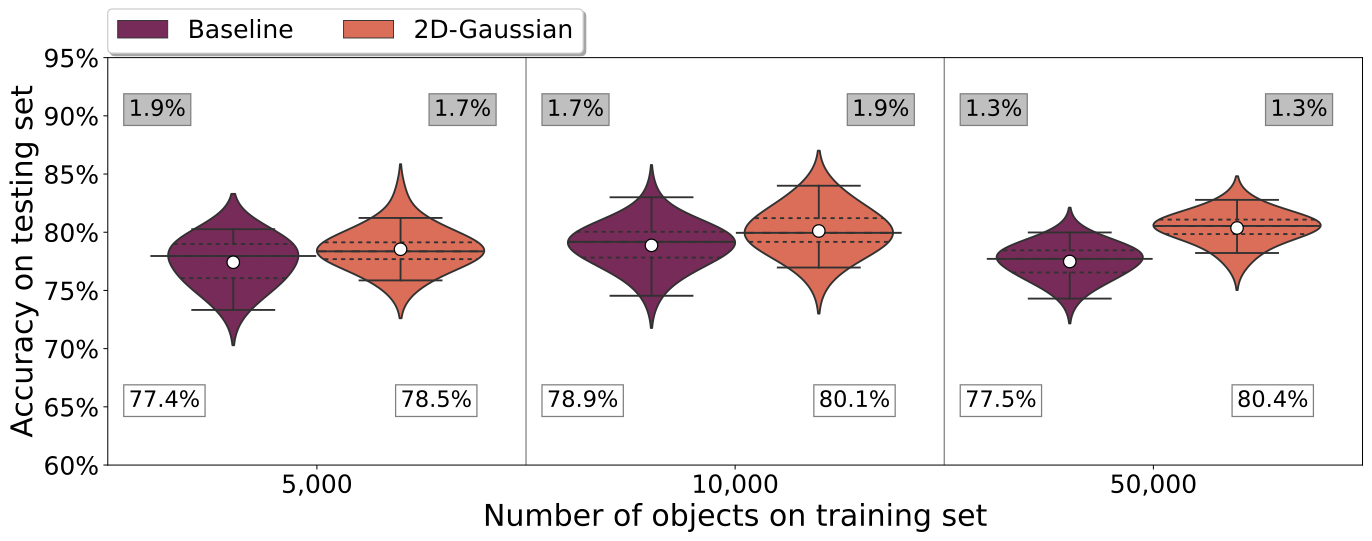}}\hspace{0mm}
\caption{Results of accuracy in baseline model and informative regularization using 2D signals. (a) provides results for 2D signals from uniform signals and (b) for 2D signals using Gaussian signals. The number of experiments and information into the boxes same as in Figure \ref{periodamplitud}. The y-axis is truncated to have a range from 60\% to 95\%.}
\label{gaussian-uniform-acc-2d}
\end{figure*}

Figure \ref{gaussian-uniform-acc-2d} and Table \ref{summaryTable} show the performance of the informative regularization adopted in our study when using 2D-signals \color{black} (amplitude and period in I band)\color{black}. Figure \ref{gaussian-uniform-acc-2d} provides violin plots, comparing accuracy metric in the testing data set, whereas Table \ref{summaryTable} gives a summary of accuracy, F1-score and AUC metrics. \color{black}

We compare eight models which are grouped as follows: (i) an MLP without regularization, which is the baseline model; (ii) the baseline model using dropout as regularization; (iii-iv) two norm-based regularization approaches over the same model ($l_1\text{-norm}$ and $l_2\text{-norm}$); and (v-viii) our four informative regularization approaches using our training procedure.  2D-Gaussian denotes signals based on bivariate Gaussian distributions and 2D-uniform denotes 2D signals from uniform distributions. \color{black}  In the 2D-Gaussian approach, the covariance matrix was estimated using a sub sample of true class objects from each sample (n =5,000, 10,000, and 50,0000); in fact, in each experiment, a different combination of objects is considered. This sub sample contains extreme values from the available objects; this filter helps to model the variability in the zone that we seek to populate (see Section \ref{2dsignalsdesign}). \color{black}

When comparing the accuracy metric in 1D signals (see  Figure \ref{periodamplitud} ) with respect to 2D signals, we can highlight a significant improvement in the models trained using 50,000 light curves. In this case, 2D-Gaussian signals improve the accuracy metric by 1.1\% from 1D-amplitude signals and 0.9\% from 1D-period signals. In the other cases (i.e., 5,000 and 10,000 training cases), 2D signals also improve the average accuracy with respect to 1D signals. Regarding standard deviation, there is no significant difference between the baseline model and models regularized by signals. 

A significant improvement in the classification performance can be noted based on the results from Table \ref{summaryTable}. For the three settings ($n=$ 5,000, 10,000 and 50,000), the informative approaches improve the baseline model. Each metric is estimated on a set of thirty experiments. The impact of our regularization proposal is noticeable, achieving the maximum increment in accuracy metric  ($\sim$3.0\%) when the maximum result for the AUC metric is considered.

\begin{table*}
\centering
\caption{Summary of results for three metrics (accuracy, F1-score and recall) for the baseline model and informative regularization. The number of experiments is the same as in Figure \ref{periodamplitud}. Early stopping is applied in all these experiments. The bold numbers represent the best strategy for model selection by each metric. $^\ast$ in bold numbers indicates a difference statistically significant with respect to the baseline model. The t-Student test was used for validating a difference in means. \color{black}}
\scriptsize
\begin{tabular}{llrrrrrrrrr}
\toprule
$n$ & Regularization  & \multicolumn{3}{c}{Accuracy} & \multicolumn{3}{c}{F1 score} & \multicolumn{3}{c}{AUC} \\       
&         &      max & median &   mean &     max & median &   mean &      max & median &  mean \\

\midrule 
5,000   & Baseline &    80.26 &  77.96 &  77.42 &   79.98 &  77.40 &  77.42 &     0.87 &   0.84 &  0.84 \\
& Dropout &    80.03 &  76.80 &  76.76 &   79.74 &  77.24 &  77.16 &     0.85 &   0.83 &  0.82 \\ 
& $l_1\text{-norm}$ ($\lambda=0.001$) &   78.41 &  71.58 &  71.57 &   75.55 &  72.65 &  72.29 &     0.84 &   0.77 &  0.77 \\
& $l_2\text{-norm}$  ($\lambda=0.001$) &    \bf{84.76} &  76.83 &  77.01 &   79.77 &  76.14 &  76.44 &     \bf{0.88} &   0.84 &  0.84 \\ 
& 1D-amplitude &    81.78 &  77.37 &  77.59 &   \bf{80.66} &  78.27 &  78.23 & \bf{0.88} &   \bf{0.85} &  \bf{0.85$^{\ast}$} \\      
& 1D-period &    80.44 &  77.35 &  77.51 &   80.08 &  78.12 &  77.80 &     0.87 &   \bf{0.85} &  \bf{0.85} \\
& 2D-Gaussian &83.20 &  \bf{78.36} &  \bf{78.54$^{\ast}$} &   80.18 &  77.77 &  77.82 &     \bf{0.88} &   \bf{0.85} &  \bf{0.85$^{\ast}$} \\ 
& 2D-uniform  & 82.26 &  77.52 &  77.81 &   80.62 &  \bf{78.40} &  \bf{78.48$^{\ast}$} &     \bf{0.88} &   \bf{0.85} &  \bf{0.85$^{\ast}$} \\
\midrule 
10,000  & Baseline &    83.01 &  79.18 &  78.87 &   81.35 &  \bf{79.22} &  79.15 &     0.88 &   0.85 &  0.85 \\       
& Dropout &    81.33 &  79.07 &  78.47 &   80.49 &  78.74 &  78.53 &     0.85 &   0.84 &  0.83 \\
& $l_1\text{-norm}$ ($\lambda=0.001$) &    73.93 &  71.87 &  72.00 &   75.47 &  73.29 &  73.12 &     0.80 &   0.78 &  0.78 \\
& $l_2\text{-norm}$  ($\lambda=0.001$) &    \bf{84.58} &  78.56 &  78.21 &   80.23 &  77.36 &  77.37 &\bf{0.89} &  \bf{0.86} & \bf{0.86$^{\ast}$} \\
& 1D-amplitude &    82.33 &  \bf{80.16} &  79.70 &   80.32 &  78.73 &  78.67 &     0.87 &   0.85 &  0.85 \\       & 1D-period &    82.68 &  79.28 &  79.52 &   80.78 &  78.70 &  78.55 &     0.87 &   0.85 &  0.85 \\       
& 2D-Gaussian &    83.99 &  79.95 &  \bf{80.11$^{\ast}$} &   \bf{81.86} &  \bf{79.22} &  \bf{79.17} &     \bf{0.89} &   \bf{0.86} &  \bf{0.86$^{\ast}$} \\
& 2D-uniform &    82.90 &  79.56 &  79.83 &   80.94 &  78.50 &  78.53 &     0.88 &   0.85 &  0.85 \\
\midrule 
50,000  & Baseline &    79.98 &  77.71 &  77.49 &   84.59 &  83.04 &  82.98 &     0.87 &   0.85 &  0.85 \\
& Dropout &    80.58 &  77.16 &  77.28 &   85.00 &  82.76 &  82.75 &     0.87 &   0.84 &  0.84 \\
& $l_1\text{-norm}$ ($\lambda=0.001$) &    72.64 &  70.91 &  70.91 &   77.52 &  75.94 &  75.94 &     0.80 &   0.77 &  0.77 \\
& $l_2\text{-norm}$  ($\lambda=0.001$) &    77.51 &  73.05 &  73.42 &   82.98 &  79.71 &  79.65 &     \bf{0.90} &   0.87 &  0.86 \\
& 1D-amplitude &    81.76 &  79.35 &  79.26 &   \bf{85.19} &  \bf{83.42} &  83.36 &     0.88 &   0.86 &  0.86 \\
& 1D-period &    81.50 &  79.46 &  79.44 &   84.53 &  83.26 &  83.33 &     0.89 &   0.87 &  0.87 \\
& 2D-Gaussian &    \bf{82.78} &  \bf{80.54} &  \bf{80.37$^{\ast}$} &   84.68 &  83.41 &  \bf{83.47$^{\ast}$} &     \bf{0.90} &   \bf{0.88} &  \bf{0.88$^{\ast}$} \\
& 2D-uniform &    80.77 &  78.94 &  79.03 &   84.71 &  83.27 &  83.13 &     0.87 &   0.86 &  0.86 \\      
\bottomrule
\end{tabular}

\label{summaryTable}
\end{table*}

Concerning the accuracy metric, when looking at experiments with $n$=5,000 in Table \ref{summaryTable}, from informative approaches, the best results were obtained by 2D-Gaussian signals; the rest of the informative regularization approaches also outperform the baseline model when the mean accuracy is considered. The improvements obtained from 2D Gaussian signals with respect to the baseline model are 2.94\%, 0.40\%, and 1.12\% for maximum, median and mean accuracy, respectively. Even more remarkable, the 2D Gaussian signals outperform all the other approaches, including $l_1\text{-norm}$, $l_2\text{-norm}$ and dropout in mean and median accuracy; 2D Gaussian approach is only outperformed by $l_2\text{-norm}$ when the maximum accuracy is considered. For the experiment using $n=10,\!000$, the maximum improvements come from $l_2\text{-norm}$ (max), 1D amplitude-based signals (mean) and  2D Gaussian signals (median). \color{black} Lastly, when 50,000 training cases are used, the biggest improvements are generated by the 2D-Gaussian signals, obtaining an increment of 2.80\% in the max metric; an increment of 2.83\% in the median metric; and an increment of 2.88\% in the mean metric. 
\color{black}

Regarding the F1-score, we also observe (see Table \ref{summaryTable})  an improvement when applying signal-based regularization. The 1D-amplitude signals provide the best result (max) with $n=5,\!000$. On the other hand, considering the mean and median metrics, 2D-uniform signals provide the best scores. For the case $n=10,\!000$ the best performance is obtained when 2D-Gaussian signals are injected, improving max accuracy by 0.5\% and median accuracy by 0.02\%; the scenario is different for $n=50,\!000$, 1D-amplitude signals achieve the best results in max and Median, and 2D-Gaussian outperforms the other alternatives when the Mean metric is prioritized. We highlight that, assessing the F1 score, informative approaches outperform traditional non-informative approaches such as dropout, $l_1\text{-norm}$ and $l_2\text{-norm}$. 

When looking at the AUC metric, we note that 2D-Gaussian signals and $l_2\text{-norm}$  obtain the best results. 2D-Gaussian signals obtain the best performance in mean AUC for the three $n$ considered. For the case $n=5,\!000$ the other informative approaches have similar performance;  for $n=10,\!000$ the two most competitive regularization methods are 2D-Gaussian and $l_2\text{-norm}$. Lastly, when $n=50,\!000$ 2D-Gaussian signals outperform the other proposals.    

These results provide evidence that our approach is a competitive regularization strategy whose main advantage is the incorporation of human knowledge by using simple synthetic data (signals). Regarding traditional regularization schemes, Dropout and $l_1\text{-norm}$ regularization methods cannot improve results according to our experiments; In fact, in most cases, these regularization methods worsen the baseline model results. $l_2\text{-norm}$ was able to improve the baseline model and, in some metrics (max accuracy and AUC), was competitive with our approach.  \color{black}

\color{black}
\subsection{Discussion}

From these empirical results, we demonstrate a positive impact of incorporating signals together with an ad-hoc training procedure. Our approach improved by up to 3\% the baseline model performance by considering several metrics (max, mean and median for accuracy, F1-score, and AUC). Moreover, the improvements were, in the majority of cases, statistically significant, and they are observed in all those metrics.   The traditional regularization behavior is expected since the main problem is related to data shift instead of overfitting. $l_1$-norm, $l_2$-norm, and dropout are methods that focus on avoiding overfitting (learn noise) but even non overfitted models decrease their performance when they are trained in a date shift scenario.

The main goal of this paper is not to perform an optimum classification of the RR Lyrae stars in our sample. Instead, we aim to demonstrate that, in the presence of significant data shifts, classifiers can be seriously misled. This problem can be ameliorated by incorporating even a small amount of expert knowledge during the training process, following the approach described in this paper. In the near future, the incorporation of additional expert knowledge, including for instance intrinsic colors and metallicities, and/or proxies such as Fourier parameters and amplitudes, may lead to further improvement in the resulting classification, beyond what was already found in this initial study of the subject.

In our experiments, we only considered 2D signals based on amplitude and period. Using a higher dimensionality ($\geq$3) for designing signals, such as including these same features from other bands (e.g., band V), could be a possibility. However, we chose to focus on the I band time series due to the limited observations per object in other bands. Another issue limiting the use of higher dimensions is that the more DRs included, the more difficult it is to generate an impact in useful feature zones.


\color{black}
\section{Conclusions}
\label{conclusions}

We propose a novel alternative for mitigating the data shift problem in tabular data by introducing rule-based knowledge into ANNs. The knowledge representation is based on characteristic ranges in astrophysically relevant parameters that must be collected from domain experts. An ad-hoc training procedure has been designed to inject this knowledge into ANNs, which considers a regularization function, masks, and a two-phase back-propagation strategy. 

Experiments show that the proposed method provides a clear improvement in three complementary metrics (accuracy, F1 score and AUC) on the shifted testing set. When 1D signals were injected into the neural net, an improvement of 2.0\% was obtained in the mean accuracy in the testing set; for the 2D signal, a statistically significant improvement of 3.0\% was achieved in the AUC metric. Hence, our method provides an alternative to perform a more reliable classification of RR Lyrae stars, even in the presence of data shift. 

Future work can consider the design and incorporation of new DRs focused on improving the classification of RR Lyrae subclasses in the presence of a data shift problem; \color{black} we could inject knowledge related to the light curve magnitude asymmetry from the available features (e.g., Rc, skew and gskew). \color{black} Additionally, this procedure could be adapted to incorporate signals into a multiclass MLP classifier. Lastly, this procedure could be applied to train a different type of neural network, for example, in a CNN where dedicated filters learn an auxiliary loss directly from light curves.    

\section*{ACKNOWLEDGEMENTS}
The authors would like to acknowledge the support from the National Agency for Research and Development (ANID), through the FONDECYT Regular project number 1180054. F. P\'erez-Galarce acknowledges support from ANID, through Scholarship Program/Doctorado Nacional/2017-21171036. P.H. acknowledges support from ANID, through FONDECYT regular 1211374.
Support for M.C. is provided by ANID's Millennium Science Initiative through grant ICN12\textunderscore 12009, awarded to the Millennium Institute of Astrophysics (MAS), and by Proyecto Basal FB210003. D. Mery acknowledges the support from ANID by Proyecto Basal FB210017, through the National Center for Artificial Intelligence CENIA.

\bibliographystyle{elsarticle-num-names}
\bibliography{references}  

\begin{thebibliography}{86}
\expandafter\ifx\csname natexlab\endcsname\relax\def\natexlab#1{#1}\fi
\providecommand{\url}[1]{\texttt{#1}}
\providecommand{\href}[2]{#2}
\providecommand{\path}[1]{#1}
\providecommand{\DOIprefix}{doi:}
\providecommand{\ArXivprefix}{arXiv:}
\providecommand{\URLprefix}{URL: }
\providecommand{\Pubmedprefix}{pmid:}
\providecommand{\doi}[1]{\href{http://dx.doi.org/#1}{\path{#1}}}
\providecommand{\Pubmed}[1]{\href{pmid:#1}{\path{#1}}}
\providecommand{\bibinfo}[2]{#2}
\ifx\xfnm\relax \def\xfnm[#1]{\unskip,\space#1}\fi
\bibitem[{Debosscher et~al.(2007)Debosscher, Sarro, Aerts, Cuypers,
  Vandenbussche, Garrido, and Solano}]{debosscher2007automated}
\bibinfo{author}{J.~Debosscher}, \bibinfo{author}{L.~Sarro},
  \bibinfo{author}{C.~Aerts}, \bibinfo{author}{J.~Cuypers},
  \bibinfo{author}{B.~Vandenbussche}, \bibinfo{author}{R.~Garrido},
  \bibinfo{author}{E.~Solano},
\newblock \bibinfo{title}{Automated supervised classification of variable
  stars},
\newblock \bibinfo{journal}{Astronomy \& Astrophysics} \bibinfo{volume}{475}
  (\bibinfo{year}{2007}) \bibinfo{pages}{1159--1183}.
\bibitem[{Debosscher et~al.(2009)Debosscher, Sarro, L{\'o}pez, Deleuil, Aerts,
  Auvergne, Baglin, Baudin, Chadid, Charpinet et~al.}]{debosscher2009automated}
\bibinfo{author}{J.~Debosscher}, \bibinfo{author}{L.~Sarro},
  \bibinfo{author}{M.~L{\'o}pez}, \bibinfo{author}{M.~Deleuil},
  \bibinfo{author}{C.~Aerts}, \bibinfo{author}{M.~Auvergne},
  \bibinfo{author}{A.~Baglin}, \bibinfo{author}{F.~Baudin},
  \bibinfo{author}{M.~Chadid}, \bibinfo{author}{S.~Charpinet}, et~al.,
\newblock \bibinfo{title}{Automated supervised classification of variable stars
  in the corot programme-method and application to the first four exoplanet
  fields},
\newblock \bibinfo{journal}{Astronomy \& Astrophysics} \bibinfo{volume}{506}
  (\bibinfo{year}{2009}) \bibinfo{pages}{519--534}.
\bibitem[{Richards et~al.(2011)Richards, Starr, Butler, Bloom, Brewer,
  Crellin-Quick, Higgins, Kennedy, and Rischard}]{richards2011machine}
\bibinfo{author}{J.~Richards}, \bibinfo{author}{D.~Starr},
  \bibinfo{author}{N.~Butler}, \bibinfo{author}{J.~Bloom},
  \bibinfo{author}{J.~Brewer}, \bibinfo{author}{A.~Crellin-Quick},
  \bibinfo{author}{J.~Higgins}, \bibinfo{author}{R.~Kennedy},
  \bibinfo{author}{M.~Rischard},
\newblock \bibinfo{title}{On machine-learned classification of variable stars
  with sparse and noisy time-series data},
\newblock \bibinfo{journal}{The Astrophysical Journal} \bibinfo{volume}{733}
  (\bibinfo{year}{2011}) \bibinfo{pages}{10}.
\bibitem[{Pichara et~al.(2012)Pichara, Protopapas, Kim, Marquette, and
  Tisserand}]{pichara2012improved}
\bibinfo{author}{K.~Pichara}, \bibinfo{author}{P.~Protopapas},
  \bibinfo{author}{D.~Kim}, \bibinfo{author}{J.~Marquette},
  \bibinfo{author}{P.~Tisserand},
\newblock \bibinfo{title}{An improved quasar detection method in eros-2 and
  macho lmc data sets},
\newblock \bibinfo{journal}{Monthly Notices of the Royal Astronomical Society}
  \bibinfo{volume}{427} (\bibinfo{year}{2012}) \bibinfo{pages}{1284--1297}.
\bibitem[{Mackenzie et~al.(2016)Mackenzie, Pichara, and
  Protopapas}]{mackenzie2016clustering}
\bibinfo{author}{C.~Mackenzie}, \bibinfo{author}{K.~Pichara},
  \bibinfo{author}{P.~Protopapas},
\newblock \bibinfo{title}{Clustering-based feature learning on variable stars},
\newblock \bibinfo{journal}{The Astrophysical Journal} \bibinfo{volume}{820}
  (\bibinfo{year}{2016}) \bibinfo{pages}{138}.
\bibitem[{Aguirre et~al.(2019)Aguirre, Pichara, and Becker}]{aguirre2018deep}
\bibinfo{author}{C.~Aguirre}, \bibinfo{author}{K.~Pichara},
  \bibinfo{author}{I.~Becker},
\newblock \bibinfo{title}{Deep multi-survey classification of variable stars},
\newblock \bibinfo{journal}{Monthly Notices of the Royal Astronomical Society}
  \bibinfo{volume}{482} (\bibinfo{year}{2019}) \bibinfo{pages}{5078--5092}.
\bibitem[{Naul et~al.(2018)Naul, Bloom, P{\'e}rez, and van~der
  Walt}]{naul2018recurrent}
\bibinfo{author}{B.~Naul}, \bibinfo{author}{J.~S. Bloom},
  \bibinfo{author}{F.~P{\'e}rez}, \bibinfo{author}{S.~van~der Walt},
\newblock \bibinfo{title}{A recurrent neural network for classification of
  unevenly sampled variable stars},
\newblock \bibinfo{journal}{Nature Astronomy} \bibinfo{volume}{2}
  (\bibinfo{year}{2018}) \bibinfo{pages}{151}.
\bibitem[{Becker et~al.(2020)Becker, Pichara, Catelan, Protopapas, Aguirre, and
  Nikzat}]{becker2020scalable}
\bibinfo{author}{I.~Becker}, \bibinfo{author}{K.~Pichara},
  \bibinfo{author}{M.~Catelan}, \bibinfo{author}{P.~Protopapas},
  \bibinfo{author}{C.~Aguirre}, \bibinfo{author}{F.~Nikzat},
\newblock \bibinfo{title}{Scalable end-to-end recurrent neural network for
  variable star classification},
\newblock \bibinfo{journal}{Monthly Notices of the Royal Astronomical Society}
  \bibinfo{volume}{493} (\bibinfo{year}{2020}) \bibinfo{pages}{2981--2995}.
\bibitem[{P{\'e}rez-Galarce et~al.(2021)P{\'e}rez-Galarce, Pichara, Huijse,
  Catelan, and Mery}]{perez2021informative}
\bibinfo{author}{F.~P{\'e}rez-Galarce}, \bibinfo{author}{K.~Pichara},
  \bibinfo{author}{P.~Huijse}, \bibinfo{author}{M.~Catelan},
  \bibinfo{author}{D.~Mery},
\newblock \bibinfo{title}{Informative bayesian model selection for rr lyrae
  star classifiers},
\newblock \bibinfo{journal}{Monthly Notices of the Royal Astronomical Society}
  \bibinfo{volume}{503} (\bibinfo{year}{2021}) \bibinfo{pages}{484--497}.
\bibitem[{Čokina et~al.(2021)Čokina, Maslej-Krešňáková, Butka, and
  Parimucha}]{COKINA2021100488}
\bibinfo{author}{M.~Čokina}, \bibinfo{author}{V.~Maslej-Krešňáková},
  \bibinfo{author}{P.~Butka}, \bibinfo{author}{S.~Parimucha},
\newblock \bibinfo{title}{Automatic classification of eclipsing binary stars
  using deep learning methods},
\newblock \bibinfo{journal}{Astronomy \& Computing} \bibinfo{volume}{36}
  (\bibinfo{year}{2021}) \bibinfo{pages}{100488}.
\bibitem[{Zhang and Bloom(2021)}]{zhang2021classification}
\bibinfo{author}{K.~Zhang}, \bibinfo{author}{J.~Bloom},
\newblock \bibinfo{title}{Classification of periodic variable stars with novel
  cyclic-permutation invariant neural networks},
\newblock \bibinfo{journal}{Monthly Notices of the Royal Astronomical Society}
  \bibinfo{volume}{505} (\bibinfo{year}{2021}) \bibinfo{pages}{515--522}.
\bibitem[{Samus et~al.(2017)Samus, Kazarovets, Durlevich, Kireeva, and
  Pastukhova}]{samus2017general}
\bibinfo{author}{N.~Samus}, \bibinfo{author}{E.~Kazarovets},
  \bibinfo{author}{O.~Durlevich}, \bibinfo{author}{N.~Kireeva},
  \bibinfo{author}{E.~Pastukhova},
\newblock \bibinfo{title}{General catalogue of variable stars: Version gcvs
  5.1},
\newblock \bibinfo{journal}{Astronomy Reports} \bibinfo{volume}{61}
  (\bibinfo{year}{2017}) \bibinfo{pages}{80--88}.
\bibitem[{Beaton et~al.(2016)Beaton, Freedman, Madore, Bono, Carlson,
  Clementini, Durbin, Garofalo, Hatt, Jang et~al.}]{beaton2016carnegie}
\bibinfo{author}{R.~Beaton}, \bibinfo{author}{W.~Freedman},
  \bibinfo{author}{B.~Madore}, \bibinfo{author}{G.~Bono},
  \bibinfo{author}{E.~Carlson}, \bibinfo{author}{G.~Clementini},
  \bibinfo{author}{M.~Durbin}, \bibinfo{author}{A.~Garofalo},
  \bibinfo{author}{D.~Hatt}, \bibinfo{author}{I.~S. Jang}, et~al.,
\newblock \bibinfo{title}{The carnegie-chicago hubble program. i. an
  independent approach to the extragalactic distance scale using only
  population ii distance indicators},
\newblock \bibinfo{journal}{The Astrophysical Journal} \bibinfo{volume}{832}
  (\bibinfo{year}{2016}) \bibinfo{pages}{210}.
\bibitem[{Pietrukowicz et~al.(2020)Pietrukowicz, Udalski, Soszynski, Skowron,
  Wrona, Szymanski, Poleski, Ulaczyk, Kozlowski, Skowron
  et~al.}]{pietrukowicz2020properties}
\bibinfo{author}{P.~Pietrukowicz}, \bibinfo{author}{A.~Udalski},
  \bibinfo{author}{I.~Soszynski}, \bibinfo{author}{D.~Skowron},
  \bibinfo{author}{M.~Wrona}, \bibinfo{author}{M.~Szymanski},
  \bibinfo{author}{R.~Poleski}, \bibinfo{author}{K.~Ulaczyk},
  \bibinfo{author}{S.~Kozlowski}, \bibinfo{author}{J.~Skowron}, et~al.,
\newblock \bibinfo{title}{Properties of the milky way's old populations based
  on photometric metallicities of the ogle rr lyrae stars},
\newblock \bibinfo{journal}{arXiv preprint arXiv:2007.05849}
  (\bibinfo{year}{2020}).
\bibitem[{D{\'e}k{\'a}ny and Grebel(2022)}]{dekany2022photometric}
\bibinfo{author}{I.~D{\'e}k{\'a}ny}, \bibinfo{author}{E.~Grebel},
\newblock \bibinfo{title}{Photometric metallicity prediction of
  fundamental-mode rr lyrae stars in the gaia optical and k s infrared wave
  bands by deep learning},
\newblock \bibinfo{journal}{The Astrophysical Journal Supplement Series}
  \bibinfo{volume}{261} (\bibinfo{year}{2022}) \bibinfo{pages}{33}.
\bibitem[{Benavente et~al.(2017)Benavente, Protopapas, and
  Pichara}]{benavente2017automatic}
\bibinfo{author}{P.~Benavente}, \bibinfo{author}{P.~Protopapas},
  \bibinfo{author}{K.~Pichara},
\newblock \bibinfo{title}{Automatic survey-invariant classification of variable
  stars},
\newblock \bibinfo{journal}{The Astrophysical Journal} \bibinfo{volume}{845}
  (\bibinfo{year}{2017}) \bibinfo{pages}{147}.
\bibitem[{Richards et~al.(2011)Richards, Starr, Brink, Miller, Bloom, Butler,
  James, Long, and Rice}]{richards2011active}
\bibinfo{author}{J.~Richards}, \bibinfo{author}{D.~Starr},
  \bibinfo{author}{H.~Brink}, \bibinfo{author}{A.~Miller},
  \bibinfo{author}{J.~Bloom}, \bibinfo{author}{N.~Butler},
  \bibinfo{author}{J.~James}, \bibinfo{author}{J.~Long},
  \bibinfo{author}{J.~Rice},
\newblock \bibinfo{title}{Active learning to overcome sample selection bias:
  application to photometric variable star classification},
\newblock \bibinfo{journal}{The Astrophysical Journal} \bibinfo{volume}{744}
  (\bibinfo{year}{2011}) \bibinfo{pages}{192}.
\bibitem[{Bloom et~al.(2012)Bloom, Richards, Nugent, Quimby, Kasliwal, Starr,
  Poznanski, Ofek, Cenko, Butler et~al.}]{bloom2012automating}
\bibinfo{author}{J.~Bloom}, \bibinfo{author}{J.~Richards},
  \bibinfo{author}{P.~Nugent}, \bibinfo{author}{R.~Quimby},
  \bibinfo{author}{M.~Kasliwal}, \bibinfo{author}{D.~Starr},
  \bibinfo{author}{D.~Poznanski}, \bibinfo{author}{E.~Ofek},
  \bibinfo{author}{S.~Cenko}, \bibinfo{author}{N.~Butler}, et~al.,
\newblock \bibinfo{title}{Automating discovery and classification of transients
  and variable stars in the synoptic survey era},
\newblock \bibinfo{journal}{Publications of the Astronomical Society of the
  Pacific} \bibinfo{volume}{124} (\bibinfo{year}{2012}) \bibinfo{pages}{1175}.
\bibitem[{Narayan et~al.(2018)Narayan, Zaidi, Soraisam, Wang, Lochner,
  Matheson, Saha, Yang, Zhao, Kececioglu et~al.}]{narayan2018machine}
\bibinfo{author}{G.~Narayan}, \bibinfo{author}{T.~Zaidi},
  \bibinfo{author}{M.~Soraisam}, \bibinfo{author}{Z.~Wang},
  \bibinfo{author}{M.~Lochner}, \bibinfo{author}{T.~Matheson},
  \bibinfo{author}{A.~Saha}, \bibinfo{author}{S.~Yang},
  \bibinfo{author}{Z.~Zhao}, \bibinfo{author}{J.~Kececioglu}, et~al.,
\newblock \bibinfo{title}{Machine-learning-based brokers for real-time
  classification of the lsst alert stream},
\newblock \bibinfo{journal}{The Astrophysical Journal Supplement Series}
  \bibinfo{volume}{236} (\bibinfo{year}{2018}) \bibinfo{pages}{9}.
\bibitem[{Masci et~al.(2014)Masci, Hoffman, Grillmair, and
  Cutri}]{masci2014automated}
\bibinfo{author}{F.~Masci}, \bibinfo{author}{D.~Hoffman},
  \bibinfo{author}{C.~Grillmair}, \bibinfo{author}{R.~Cutri},
\newblock \bibinfo{title}{Automated classification of periodic variable stars
  detected by the wide-field infrared survey explorer},
\newblock \bibinfo{journal}{The Astronomical Journal} \bibinfo{volume}{148}
  (\bibinfo{year}{2014}) \bibinfo{pages}{21}.
\bibitem[{{Rebbapragada} et~al.(2015){Rebbapragada}, {Bue}, and
  {Wozniak}}]{Rebbapragada}
\bibinfo{author}{U.~{Rebbapragada}}, \bibinfo{author}{B.~{Bue}},
  \bibinfo{author}{P.~{Wozniak}},
\newblock \bibinfo{title}{{Time-domain Surveys and Data Shift: Case Study at
  the intermediate Palomar Transient Factory}},
\newblock in: \bibinfo{booktitle}{American Astronomical Society Meeting
  Abstracts \#225}, volume \bibinfo{volume}{225} of
  \textit{\bibinfo{series}{American Astronomical Society Meeting Abstracts}},
  \bibinfo{year}{2015}, p. \bibinfo{pages}{434.02}.
\bibitem[{Quiñonero-Candela et~al.(2009)Quiñonero-Candela, Sugiyama,
  Schwaighofer, and Lawrence}]{candela2009dataset}
\bibinfo{author}{J.~Quiñonero-Candela}, \bibinfo{author}{M.~Sugiyama},
  \bibinfo{author}{A.~Schwaighofer}, \bibinfo{author}{N.~Lawrence},
  \bibinfo{title}{Dataset shift in machine learning}, \bibinfo{publisher}{The
  MIT Press, Cambridge, Massachusetts}, \bibinfo{year}{2009}.
\bibitem[{Clemmensen and Kj{\ae}rsgaard(2022)}]{clemmensen2022data}
\bibinfo{author}{L.~H. Clemmensen}, \bibinfo{author}{R.~D. Kj{\ae}rsgaard},
\newblock \bibinfo{title}{Data representativity for machine learning and ai
  systems},
\newblock \bibinfo{journal}{arXiv preprint arXiv:2203.04706}
  (\bibinfo{year}{2022}).
\bibitem[{Chawla et~al.(2004)Chawla, Japkowicz, and Kotcz}]{chawla2004special}
\bibinfo{author}{N.~V. Chawla}, \bibinfo{author}{N.~Japkowicz},
  \bibinfo{author}{A.~Kotcz},
\newblock \bibinfo{title}{Special issue on learning from imbalanced data sets},
\newblock \bibinfo{journal}{ACM SIGKDD explorations newsletter}
  \bibinfo{volume}{6} (\bibinfo{year}{2004}) \bibinfo{pages}{1--6}.
\bibitem[{Cabrera et~al.(2014)Cabrera, Miller, and
  Schneider}]{cabrera2014systematic}
\bibinfo{author}{G.~F. Cabrera}, \bibinfo{author}{C.~J. Miller},
  \bibinfo{author}{J.~Schneider},
\newblock \bibinfo{title}{Systematic labeling bias: De-biasing where everyone
  is wrong},
\newblock in: \bibinfo{booktitle}{22nd International Conference on Pattern
  Recognition}, \bibinfo{organization}{IEEE}, \bibinfo{year}{2014}, pp.
  \bibinfo{pages}{4417--4422}.
\bibitem[{Svozil et~al.(1997)Svozil, Kvasnicka, and
  Pospichal}]{svozil1997introduction}
\bibinfo{author}{D.~Svozil}, \bibinfo{author}{V.~Kvasnicka},
  \bibinfo{author}{J.~Pospichal},
\newblock \bibinfo{title}{Introduction to multi-layer feed-forward neural
  networks},
\newblock \bibinfo{journal}{Chemometrics and intelligent laboratory systems}
  \bibinfo{volume}{39} (\bibinfo{year}{1997}) \bibinfo{pages}{43--62}.
\bibitem[{Hornik et~al.(1989)Hornik, Stinchcombe, and
  White}]{hornik1989multilayer}
\bibinfo{author}{K.~Hornik}, \bibinfo{author}{M.~Stinchcombe},
  \bibinfo{author}{H.~White},
\newblock \bibinfo{title}{Multilayer feedforward networks are universal
  approximators},
\newblock \bibinfo{journal}{Neural networks} \bibinfo{volume}{2}
  (\bibinfo{year}{1989}) \bibinfo{pages}{359--366}.
\bibitem[{Nun et~al.(2014)Nun, Pichara, Protopapas, and
  Kim}]{nun2014supervised}
\bibinfo{author}{I.~Nun}, \bibinfo{author}{K.~Pichara},
  \bibinfo{author}{P.~Protopapas}, \bibinfo{author}{D.-W. Kim},
\newblock \bibinfo{title}{Supervised detection of anomalous light curves in
  massive astronomical catalogs},
\newblock \bibinfo{journal}{The Astrophysical Journal} \bibinfo{volume}{793}
  (\bibinfo{year}{2014}) \bibinfo{pages}{23}.
\bibitem[{Nun et~al.(2015)Nun, Protopapas, Sim, Zhu, Dave, Castro, and
  Pichara}]{nun2015fats}
\bibinfo{author}{I.~Nun}, \bibinfo{author}{P.~Protopapas},
  \bibinfo{author}{B.~Sim}, \bibinfo{author}{M.~Zhu},
  \bibinfo{author}{R.~Dave}, \bibinfo{author}{N.~Castro},
  \bibinfo{author}{K.~Pichara},
\newblock \bibinfo{title}{Fats: Feature analysis for time series},
\newblock \bibinfo{journal}{arXiv preprint arXiv:1506.00010}
  (\bibinfo{year}{2015}).
\bibitem[{Kim and Bailer-Jones(2016)}]{kim2016package}
\bibinfo{author}{D.~Kim}, \bibinfo{author}{C.~Bailer-Jones},
\newblock \bibinfo{title}{A package for the automated classification of
  periodic variable stars},
\newblock \bibinfo{journal}{Astronomy \& Astrophysics} \bibinfo{volume}{587}
  (\bibinfo{year}{2016}) \bibinfo{pages}{A18}.
\bibitem[{{Cabral} et~al.(2018){Cabral}, {S{\'a}nchez}, {Ramos}, {Gurovich},
  {Granitto}, and {Vanderplas}}]{feets}
\bibinfo{author}{J.~B. {Cabral}}, \bibinfo{author}{B.~{S{\'a}nchez}},
  \bibinfo{author}{F.~{Ramos}}, \bibinfo{author}{S.~{Gurovich}},
  \bibinfo{author}{P.~M. {Granitto}}, \bibinfo{author}{J.~{Vanderplas}},
\newblock \bibinfo{title}{{From FATS to feets: Further improvements to an
  astronomical feature extraction tool based on machine learning}},
\newblock \bibinfo{journal}{Astronomy \& Computing} \bibinfo{volume}{25}
  (\bibinfo{year}{2018}) \bibinfo{pages}{213--220}.
\bibitem[{Pichara et~al.(2016)Pichara, Protopapas, and
  Le{\'o}n}]{pichara2016meta}
\bibinfo{author}{K.~Pichara}, \bibinfo{author}{P.~Protopapas},
  \bibinfo{author}{D.~Le{\'o}n},
\newblock \bibinfo{title}{Meta-classification for variable stars},
\newblock \bibinfo{journal}{The Astrophysical Journal} \bibinfo{volume}{819}
  (\bibinfo{year}{2016}) \bibinfo{pages}{18}.
\bibitem[{Hartman and Bakos(2016)}]{hartman2016vartools}
\bibinfo{author}{J.~D. Hartman}, \bibinfo{author}{G.~{\'A}. Bakos},
\newblock \bibinfo{title}{Vartools: A program for analyzing astronomical
  time-series data},
\newblock \bibinfo{journal}{Astronomy \& Computing} \bibinfo{volume}{17}
  (\bibinfo{year}{2016}) \bibinfo{pages}{1--72}.
\bibitem[{VanderPlas(2016)}]{vanderplas2016gatspy}
\bibinfo{author}{J.~VanderPlas},
\newblock \bibinfo{title}{gatspy: General tools for astronomical time series in
  python},
\newblock \bibinfo{journal}{Astrophysics Source Code Library}
  (\bibinfo{year}{2016}) \bibinfo{pages}{ascl--1610}.
\bibitem[{Naul et~al.(2016)Naul, van~der Walt, Crellin-Quick, Bloom, and
  P{\'e}rez}]{naul2016cesium}
\bibinfo{author}{B.~Naul}, \bibinfo{author}{S.~van~der Walt},
  \bibinfo{author}{A.~Crellin-Quick}, \bibinfo{author}{J.~S. Bloom},
  \bibinfo{author}{F.~P{\'e}rez},
\newblock \bibinfo{title}{cesium: Open-source platform for time-series
  inference},
\newblock \bibinfo{journal}{arXiv preprint arXiv:1609.04504}
  (\bibinfo{year}{2016}).
\bibitem[{Barbary et~al.(2016)Barbary, Barclay, Biswas, Craig, Feindt, Friesen,
  Goldstein, Jha, Rodney, Sofiatti et~al.}]{barbary2016sncosmo}
\bibinfo{author}{K.~Barbary}, \bibinfo{author}{T.~Barclay},
  \bibinfo{author}{R.~Biswas}, \bibinfo{author}{M.~Craig},
  \bibinfo{author}{U.~Feindt}, \bibinfo{author}{B.~Friesen},
  \bibinfo{author}{D.~Goldstein}, \bibinfo{author}{S.~Jha},
  \bibinfo{author}{S.~Rodney}, \bibinfo{author}{C.~Sofiatti}, et~al.,
\newblock \bibinfo{title}{Sncosmo: Python library for supernova cosmology},
\newblock \bibinfo{journal}{Astrophysics Source Code Library}
  (\bibinfo{year}{2016}) \bibinfo{pages}{ascl--1611}.
\bibitem[{Wright et~al.(2010)Wright, Eisenhardt, Mainzer, Ressler, Cutri,
  Jarrett, Kirkpatrick, Padgett, McMillan, Skrutskie et~al.}]{wright2010wide}
\bibinfo{author}{E.~Wright}, \bibinfo{author}{P.~Eisenhardt},
  \bibinfo{author}{A.~Mainzer}, \bibinfo{author}{M.~Ressler},
  \bibinfo{author}{R.~Cutri}, \bibinfo{author}{T.~Jarrett},
  \bibinfo{author}{D.~Kirkpatrick}, \bibinfo{author}{D.~Padgett},
  \bibinfo{author}{R.~McMillan}, \bibinfo{author}{M.~Skrutskie}, et~al.,
\newblock \bibinfo{title}{The wide-field infrared survey explorer (wise):
  mission description and initial on-orbit performance},
\newblock \bibinfo{journal}{The Astronomical Journal} \bibinfo{volume}{140}
  (\bibinfo{year}{2010}) \bibinfo{pages}{1868}.
\bibitem[{Settles(2009)}]{settles2009active}
\bibinfo{author}{B.~Settles}, \bibinfo{title}{Active learning literature
  survey}, \bibinfo{publisher}{University of Wisconsin-Madison Department of
  Computer Sciences}, \bibinfo{year}{2009}.
\bibitem[{Carrasco-Davis et~al.(2019)Carrasco-Davis, Cabrera-Vives,
  F{\"o}rster, Est{\'e}vez, Huijse, Protopapas, Reyes, Mart{\'\i}nez-Palomera,
  and Donoso}]{carrasco2018deep}
\bibinfo{author}{R.~Carrasco-Davis}, \bibinfo{author}{G.~Cabrera-Vives},
  \bibinfo{author}{F.~F{\"o}rster}, \bibinfo{author}{P.~A. Est{\'e}vez},
  \bibinfo{author}{P.~Huijse}, \bibinfo{author}{P.~Protopapas},
  \bibinfo{author}{I.~Reyes}, \bibinfo{author}{J.~Mart{\'\i}nez-Palomera},
  \bibinfo{author}{C.~Donoso},
\newblock \bibinfo{title}{Deep learning for image sequence classification of
  astronomical events},
\newblock \bibinfo{journal}{Publications of the Astronomical Society of the
  Pacific} \bibinfo{volume}{131} (\bibinfo{year}{2019})
  \bibinfo{pages}{108006}.
\bibitem[{Jamal and Bloom(2020)}]{jamal2020neural}
\bibinfo{author}{S.~Jamal}, \bibinfo{author}{J.~Bloom},
\newblock \bibinfo{title}{On neural architectures for astronomical time-series
  classification with application to variable stars},
\newblock \bibinfo{journal}{The Astrophysical Journal Supplement Series}
  \bibinfo{volume}{250} (\bibinfo{year}{2020}) \bibinfo{pages}{30}.
\bibitem[{Udalski et~al.(2008)Udalski, Szymanski, Soszynski, and
  Poleski}]{udalski2008optical}
\bibinfo{author}{A.~Udalski}, \bibinfo{author}{M.~Szymanski},
  \bibinfo{author}{I.~Soszynski}, \bibinfo{author}{R.~Poleski},
\newblock \bibinfo{title}{The optical gravitational lensing experiment. final
  reductions of the ogle-iii data},
\newblock \bibinfo{journal}{arXiv preprint arXiv:0807.3884}
  (\bibinfo{year}{2008}).
\bibitem[{Alcock et~al.(1997)Alcock, Allsman, Alves, Axelrod, Becker, Bennett,
  Cook, Freeman, Griest, Guern et~al.}]{alcock1997macho}
\bibinfo{author}{C.~Alcock}, \bibinfo{author}{R.~Allsman},
  \bibinfo{author}{D.~Alves}, \bibinfo{author}{T.~Axelrod},
  \bibinfo{author}{A.~Becker}, \bibinfo{author}{D.~Bennett},
  \bibinfo{author}{K.~Cook}, \bibinfo{author}{K.~Freeman},
  \bibinfo{author}{K.~Griest}, \bibinfo{author}{J.~Guern}, et~al.,
\newblock \bibinfo{title}{The macho project large magellanic cloud microlensing
  results from the first two years and the nature of the galactic dark halo},
\newblock \bibinfo{journal}{The Astrophysical Journal} \bibinfo{volume}{486}
  (\bibinfo{year}{1997}) \bibinfo{pages}{697}.
\bibitem[{Jayasinghe et~al.(2019)Jayasinghe, Stanek, Kochanek, Shappee,
  Holoien, Thompson, Prieto, Dong, Pawlak, Pejcha et~al.}]{jayasinghe2019asas}
\bibinfo{author}{T.~Jayasinghe}, \bibinfo{author}{K.~Stanek},
  \bibinfo{author}{C.~Kochanek}, \bibinfo{author}{B.~Shappee},
  \bibinfo{author}{T.~W. Holoien}, \bibinfo{author}{T.~A. Thompson},
  \bibinfo{author}{J.~Prieto}, \bibinfo{author}{S.~Dong},
  \bibinfo{author}{M.~Pawlak}, \bibinfo{author}{O.~Pejcha}, et~al.,
\newblock \bibinfo{title}{The asas-sn catalogue of variable stars--ii. uniform
  classification of 412 000 known variables},
\newblock \bibinfo{journal}{Monthly Notices of the Royal Astronomical Society}
  \bibinfo{volume}{486} (\bibinfo{year}{2019}) \bibinfo{pages}{1907--1943}.
\bibitem[{Catelan and Smith(2015)}]{catelan2014pulsating}
\bibinfo{author}{M.~Catelan}, \bibinfo{author}{H.~Smith},
  \bibinfo{title}{Pulsating Stars}, \bibinfo{publisher}{Wiley-VCH, Weinheim},
  \bibinfo{year}{2015}.
\bibitem[{Gran et~al.(2016)Gran, Minniti, Saito, Zoccali, Gonzalez, Navarrete,
  Catelan, Ramos, Elorrieta, Eyheramendy et~al.}]{gran2016mapping}
\bibinfo{author}{F.~Gran}, \bibinfo{author}{D.~Minniti},
  \bibinfo{author}{R.~Saito}, \bibinfo{author}{M.~Zoccali},
  \bibinfo{author}{O.~Gonzalez}, \bibinfo{author}{C.~Navarrete},
  \bibinfo{author}{M.~Catelan}, \bibinfo{author}{R.~C. Ramos},
  \bibinfo{author}{F.~Elorrieta}, \bibinfo{author}{S.~Eyheramendy}, et~al.,
\newblock \bibinfo{title}{Mapping the outer bulge with rrab stars from the vvv
  survey},
\newblock \bibinfo{journal}{Astronomy \& Astrophysics} \bibinfo{volume}{591}
  (\bibinfo{year}{2016}) \bibinfo{pages}{A145}.
\bibitem[{Minniti et~al.(2010)Minniti, Lucas, Emerson, Saito, Hempel,
  Pietrukowicz, Ahumada, Alonso, Alonso-Garcia, Arias
  et~al.}]{minniti2010vista}
\bibinfo{author}{D.~Minniti}, \bibinfo{author}{P.~Lucas},
  \bibinfo{author}{J.~Emerson}, \bibinfo{author}{R.~Saito},
  \bibinfo{author}{M.~Hempel}, \bibinfo{author}{P.~Pietrukowicz},
  \bibinfo{author}{A.~Ahumada}, \bibinfo{author}{M.~Alonso},
  \bibinfo{author}{J.~Alonso-Garcia}, \bibinfo{author}{J.~I. Arias}, et~al.,
\newblock \bibinfo{title}{Vista variables in the via lactea (vvv): The public
  eso near-ir variability survey of the milky way},
\newblock \bibinfo{journal}{New Astronomy} \bibinfo{volume}{15}
  (\bibinfo{year}{2010}) \bibinfo{pages}{433--443}.
\bibitem[{Elorrieta et~al.(2016)Elorrieta, Eyheramendy, Jord{\'a}n,
  D{\'e}k{\'a}ny, Catelan, Angeloni, Alonso-Garc{\'\i}a, Contreras-Ramos, Gran,
  Hajdu et~al.}]{elorrieta2016machine}
\bibinfo{author}{F.~Elorrieta}, \bibinfo{author}{S.~Eyheramendy},
  \bibinfo{author}{A.~Jord{\'a}n}, \bibinfo{author}{I.~D{\'e}k{\'a}ny},
  \bibinfo{author}{M.~Catelan}, \bibinfo{author}{R.~Angeloni},
  \bibinfo{author}{J.~Alonso-Garc{\'\i}a},
  \bibinfo{author}{R.~Contreras-Ramos}, \bibinfo{author}{F.~Gran},
  \bibinfo{author}{G.~Hajdu}, et~al.,
\newblock \bibinfo{title}{A machine learned classifier for rr lyrae in the vvv
  survey},
\newblock \bibinfo{journal}{Astronomy \& Astrophysics} \bibinfo{volume}{595}
  (\bibinfo{year}{2016}) \bibinfo{pages}{A82}.
\bibitem[{Sesar et~al.(2017)Sesar, Hernitschek, Mitrovi{\'c}, Ivezi{\'c}, Rix,
  Cohen, Bernard, Grebel, Martin, Schlafly et~al.}]{sesar2017machine}
\bibinfo{author}{B.~Sesar}, \bibinfo{author}{N.~Hernitschek},
  \bibinfo{author}{S.~Mitrovi{\'c}}, \bibinfo{author}{{\v{Z}}.~Ivezi{\'c}},
  \bibinfo{author}{H.-W. Rix}, \bibinfo{author}{J.~G. Cohen},
  \bibinfo{author}{E.~J. Bernard}, \bibinfo{author}{E.~K. Grebel},
  \bibinfo{author}{N.~F. Martin}, \bibinfo{author}{E.~F. Schlafly}, et~al.,
\newblock \bibinfo{title}{Machine-learned identification of rr lyrae stars from
  sparse, multi-band data: the ps1 sample},
\newblock \bibinfo{journal}{The Astronomical Journal} \bibinfo{volume}{153}
  (\bibinfo{year}{2017}) \bibinfo{pages}{204}.
\bibitem[{Kaiser et~al.(2010)Kaiser, Burgett, Chambers, and
  Denneau}]{kaiser2010ground}
\bibinfo{author}{N.~Kaiser}, \bibinfo{author}{W.~Burgett},
  \bibinfo{author}{K.~Chambers}, \bibinfo{author}{L.~Denneau},
\newblock \bibinfo{title}{Ground-based and airborne telescopes iii},
\newblock in: \bibinfo{booktitle}{Proc. SPIE}, volume \bibinfo{volume}{7733},
  \bibinfo{year}{2010}, p. \bibinfo{pages}{77330E}.
\bibitem[{D{\'e}k{\'a}ny and Grebel(2020)}]{dekany2020near}
\bibinfo{author}{I.~D{\'e}k{\'a}ny}, \bibinfo{author}{E.~K. Grebel},
\newblock \bibinfo{title}{Near-infrared search for fundamental-mode rr lyrae
  stars toward the inner bulge by deep learning},
\newblock \bibinfo{journal}{The Astrophysical Journal} \bibinfo{volume}{898}
  (\bibinfo{year}{2020}) \bibinfo{pages}{46}.
\bibitem[{Deng et~al.(2020)Deng, Ji, Rainey, Zhang, and
  Lu}]{deng2020integrating}
\bibinfo{author}{C.~Deng}, \bibinfo{author}{X.~Ji},
  \bibinfo{author}{C.~Rainey}, \bibinfo{author}{J.~Zhang},
  \bibinfo{author}{W.~Lu},
\newblock \bibinfo{title}{Integrating machine learning with human knowledge},
\newblock \bibinfo{journal}{Iscience}  (\bibinfo{year}{2020})
  \bibinfo{pages}{101656}.
\bibitem[{von Rueden et~al.(2019)von Rueden, Mayer, Beckh, Georgiev,
  Giesselbach, Heese, Kirsch, Pfrommer, Pick, Ramamurthy
  et~al.}]{von2019informed}
\bibinfo{author}{L.~von Rueden}, \bibinfo{author}{S.~Mayer},
  \bibinfo{author}{K.~Beckh}, \bibinfo{author}{B.~Georgiev},
  \bibinfo{author}{S.~Giesselbach}, \bibinfo{author}{R.~Heese},
  \bibinfo{author}{B.~Kirsch}, \bibinfo{author}{J.~Pfrommer},
  \bibinfo{author}{A.~Pick}, \bibinfo{author}{R.~Ramamurthy}, et~al.,
\newblock \bibinfo{title}{Informed machine learning--a taxonomy and survey of
  integrating knowledge into learning systems},
\newblock \bibinfo{journal}{arXiv preprint arXiv:1903.12394}
  (\bibinfo{year}{2019}).
\bibitem[{Borghesi et~al.(2020)Borghesi, Baldo, and
  Milano}]{borghesi2020improving}
\bibinfo{author}{A.~Borghesi}, \bibinfo{author}{F.~Baldo},
  \bibinfo{author}{M.~Milano},
\newblock \bibinfo{title}{Improving deep learning models via constraint-based
  domain knowledge: a brief survey},
\newblock \bibinfo{journal}{arXiv preprint arXiv:2005.10691}
  (\bibinfo{year}{2020}).
\bibitem[{Battaglia et~al.(2018)Battaglia, Hamrick, Bapst, Sanchez-Gonzalez,
  Zambaldi, Malinowski, Tacchetti, Raposo, Santoro, Faulkner
  et~al.}]{battaglia2018relational}
\bibinfo{author}{P.~Battaglia}, \bibinfo{author}{J.~Hamrick},
  \bibinfo{author}{V.~Bapst}, \bibinfo{author}{A.~Sanchez-Gonzalez},
  \bibinfo{author}{V.~Zambaldi}, \bibinfo{author}{M.~Malinowski},
  \bibinfo{author}{A.~Tacchetti}, \bibinfo{author}{D.~Raposo},
  \bibinfo{author}{A.~Santoro}, \bibinfo{author}{R.~Faulkner}, et~al.,
\newblock \bibinfo{title}{Relational inductive biases, deep learning, and graph
  networks},
\newblock \bibinfo{journal}{arXiv preprint arXiv:1806.01261}
  (\bibinfo{year}{2018}).
\bibitem[{Blanton and Roweis(2007)}]{blanton2007k}
\bibinfo{author}{M.~Blanton}, \bibinfo{author}{S.~Roweis},
\newblock \bibinfo{title}{K-corrections and filter transformations in the
  ultraviolet, optical, and near-infrared},
\newblock \bibinfo{journal}{The Astronomical Journal} \bibinfo{volume}{133}
  (\bibinfo{year}{2007}) \bibinfo{pages}{734}.
\bibitem[{Sravan et~al.(2020)Sravan, Milisavljevic, Reynolds, Lentner, and
  Linvill}]{sravan2020real}
\bibinfo{author}{N.~Sravan}, \bibinfo{author}{D.~Milisavljevic},
  \bibinfo{author}{J.~M. Reynolds}, \bibinfo{author}{G.~Lentner},
  \bibinfo{author}{M.~Linvill},
\newblock \bibinfo{title}{Real-time, value-driven data augmentation in the era
  of lsst},
\newblock \bibinfo{journal}{The Astrophysical Journal} \bibinfo{volume}{893}
  (\bibinfo{year}{2020}) \bibinfo{pages}{127}.
\bibitem[{Shorten and Khoshgoftaar(2019)}]{shorten2019survey}
\bibinfo{author}{C.~Shorten}, \bibinfo{author}{T.~M. Khoshgoftaar},
\newblock \bibinfo{title}{A survey on image data augmentation for deep
  learning},
\newblock \bibinfo{journal}{Journal of big data} \bibinfo{volume}{6}
  (\bibinfo{year}{2019}) \bibinfo{pages}{1--48}.
\bibitem[{Castro et~al.(2017)Castro, Protopapas, and
  Pichara}]{castro2017uncertain}
\bibinfo{author}{N.~Castro}, \bibinfo{author}{P.~Protopapas},
  \bibinfo{author}{K.~Pichara},
\newblock \bibinfo{title}{Uncertain classification of variable stars: Handling
  observational gaps and noise},
\newblock \bibinfo{journal}{The Astronomical Journal} \bibinfo{volume}{155}
  (\bibinfo{year}{2017}) \bibinfo{pages}{16}.
\bibitem[{Hanson et~al.(2014)Hanson, Branscum, Johnson
  et~al.}]{hanson2014informative}
\bibinfo{author}{T.~Hanson}, \bibinfo{author}{A.~J. Branscum},
  \bibinfo{author}{W.~Johnson}, et~al.,
\newblock \bibinfo{title}{Informative $ g $-priors for logistic regression},
\newblock \bibinfo{journal}{Bayesian Analysis} \bibinfo{volume}{9}
  (\bibinfo{year}{2014}) \bibinfo{pages}{597--612}.
\bibitem[{Fortuin(2022)}]{fortuin2022priors}
\bibinfo{author}{V.~Fortuin},
\newblock \bibinfo{title}{Priors in bayesian deep learning: A review},
\newblock \bibinfo{journal}{International Statistical Review}
  (\bibinfo{year}{2022}).
\bibitem[{Bollacker et~al.(2007)Bollacker, Cook, and
  Tufts}]{bollacker2007freebase}
\bibinfo{author}{K.~Bollacker}, \bibinfo{author}{R.~Cook},
  \bibinfo{author}{P.~Tufts},
\newblock \bibinfo{title}{Freebase: A shared database of structured general
  human knowledge},
\newblock in: \bibinfo{booktitle}{AAAI}, volume~\bibinfo{volume}{7},
  \bibinfo{year}{2007}, pp. \bibinfo{pages}{1962--1963}.
\bibitem[{Auer et~al.(2007)Auer, Bizer, Kobilarov, Lehmann, Cyganiak, and
  Ives}]{auer2007dbpedia}
\bibinfo{author}{S.~Auer}, \bibinfo{author}{C.~Bizer},
  \bibinfo{author}{G.~Kobilarov}, \bibinfo{author}{J.~Lehmann},
  \bibinfo{author}{R.~Cyganiak}, \bibinfo{author}{Z.~Ives},
\newblock \bibinfo{title}{Dbpedia: A nucleus for a web of open data},
\newblock in: \bibinfo{booktitle}{The semantic web},
  \bibinfo{publisher}{Springer}, \bibinfo{year}{2007}, pp.
  \bibinfo{pages}{722--735}.
\bibitem[{Suchanek et~al.(2007)Suchanek, Kasneci, and
  Weikum}]{suchanek2007yago}
\bibinfo{author}{F.~Suchanek}, \bibinfo{author}{G.~Kasneci},
  \bibinfo{author}{G.~Weikum},
\newblock \bibinfo{title}{Yago: a core of semantic knowledge},
\newblock in: \bibinfo{booktitle}{Proceedings of the 16th international
  conference on World Wide Web}, \bibinfo{year}{2007}, pp.
  \bibinfo{pages}{697--706}.
\bibitem[{Kafle et~al.(2020)Kafle, de~Silva, and Dou}]{kafle2020overview}
\bibinfo{author}{S.~Kafle}, \bibinfo{author}{N.~de~Silva},
  \bibinfo{author}{D.~Dou},
\newblock \bibinfo{title}{An overview of utilizing knowledge bases in neural
  networks for question answering},
\newblock \bibinfo{journal}{Information Systems Frontiers} \bibinfo{volume}{22}
  (\bibinfo{year}{2020}) \bibinfo{pages}{1095--1111}.
\bibitem[{Atzmueller and Sternberg(2017)}]{atzmueller2017mixed}
\bibinfo{author}{M.~Atzmueller}, \bibinfo{author}{E.~Sternberg},
\newblock \bibinfo{title}{Mixed-initiative feature engineering using knowledge
  graphs},
\newblock in: \bibinfo{booktitle}{K-CAP}, \bibinfo{year}{2017}, pp.
  \bibinfo{pages}{1--4}.
\bibitem[{Raissi et~al.(2019)Raissi, Perdikaris, and
  Karniadakis}]{raissi2019physics}
\bibinfo{author}{M.~Raissi}, \bibinfo{author}{P.~Perdikaris},
  \bibinfo{author}{G.~E. Karniadakis},
\newblock \bibinfo{title}{Physics-informed neural networks: A deep learning
  framework for solving forward and inverse problems involving nonlinear
  partial differential equations},
\newblock \bibinfo{journal}{Journal of Computational Physics}
  \bibinfo{volume}{378} (\bibinfo{year}{2019}) \bibinfo{pages}{686--707}.
\bibitem[{Goodfellow et~al.(2016)Goodfellow, Bengio, and
  Courville}]{goodfellow2016deep}
\bibinfo{author}{I.~Goodfellow}, \bibinfo{author}{Y.~Bengio},
  \bibinfo{author}{A.~Courville}, \bibinfo{title}{Deep learning},
  \bibinfo{publisher}{MIT press}, \bibinfo{year}{2016}.
\bibitem[{Neyshabur et~al.(2014)Neyshabur, Tomioka, and
  Srebro}]{neyshabur2014search}
\bibinfo{author}{B.~Neyshabur}, \bibinfo{author}{R.~Tomioka},
  \bibinfo{author}{N.~Srebro},
\newblock \bibinfo{title}{In search of the real inductive bias: On the role of
  implicit regularization in deep learning},
\newblock \bibinfo{journal}{arXiv preprint arXiv:1412.6614}
  (\bibinfo{year}{2014}).
\bibitem[{Leimkuhler et~al.(2020)Leimkuhler, Pouchon, Vlaar, and
  Storkey}]{leimkuhler2020constraint}
\bibinfo{author}{B.~Leimkuhler}, \bibinfo{author}{T.~Pouchon},
  \bibinfo{author}{T.~Vlaar}, \bibinfo{author}{A.~Storkey},
\newblock \bibinfo{title}{Constraint-based regularization of neural networks},
\newblock \bibinfo{journal}{arXiv preprint arXiv:2006.10114}
  (\bibinfo{year}{2020}).
\bibitem[{Tibshirani(1996)}]{tibshirani1996regression}
\bibinfo{author}{R.~Tibshirani},
\newblock \bibinfo{title}{Regression shrinkage and selection via the lasso},
\newblock \bibinfo{journal}{Journal of the Royal Statistical Society: Series B}
  \bibinfo{volume}{58} (\bibinfo{year}{1996}) \bibinfo{pages}{267--288}.
\bibitem[{Geman et~al.(1992)Geman, Bienenstock, and Doursat}]{geman1992neural}
\bibinfo{author}{S.~Geman}, \bibinfo{author}{E.~Bienenstock},
  \bibinfo{author}{R.~Doursat},
\newblock \bibinfo{title}{Neural networks and the bias/variance dilemma},
\newblock \bibinfo{journal}{Neural computation} \bibinfo{volume}{4}
  (\bibinfo{year}{1992}) \bibinfo{pages}{1--58}.
\bibitem[{Hoerl and Kennard(2000)}]{ridge}
\bibinfo{author}{A.~E. Hoerl}, \bibinfo{author}{R.~W. Kennard},
\newblock \bibinfo{title}{Ridge regression: Biased estimation for nonorthogonal
  problems},
\newblock \bibinfo{journal}{Technometrics} \bibinfo{volume}{42}
  (\bibinfo{year}{2000}) \bibinfo{pages}{80--86}.
\bibitem[{Zou and Hastie(2005)}]{zou2005regularization}
\bibinfo{author}{H.~Zou}, \bibinfo{author}{T.~Hastie},
\newblock \bibinfo{title}{Regularization and variable selection via the elastic
  net},
\newblock \bibinfo{journal}{Journal of the Royal Statistical Society: Series B}
  \bibinfo{volume}{67} (\bibinfo{year}{2005}) \bibinfo{pages}{301--320}.
\bibitem[{Kim et~al.(2021)Kim, Ko, and Seo}]{kim2021novel}
\bibinfo{author}{B.~Kim}, \bibinfo{author}{Y.~Ko}, \bibinfo{author}{J.~Seo},
\newblock \bibinfo{title}{Novel regularization method for the class imbalance
  problem},
\newblock \bibinfo{journal}{Expert Systems with Applications}
  (\bibinfo{year}{2021}) \bibinfo{pages}{115974}.
\bibitem[{Baldassarre et~al.(2012)Baldassarre, Mourao-Miranda, and
  Pontil}]{baldassarre2012structure}
\bibinfo{author}{L.~Baldassarre}, \bibinfo{author}{J.~Mourao-Miranda},
  \bibinfo{author}{M.~Pontil},
\newblock \bibinfo{title}{Structured sparsity models for brain decoding from
  fmri data},
\newblock in: \bibinfo{booktitle}{Second International Workshop on Pattern
  Recognition in NeuroImaging}, \bibinfo{organization}{IEEE},
  \bibinfo{year}{2012}, pp. \bibinfo{pages}{5--8}.
\bibitem[{Chambolle(2004)}]{chambolle2004algorithm}
\bibinfo{author}{A.~Chambolle},
\newblock \bibinfo{title}{An algorithm for total variation minimization and
  applications},
\newblock \bibinfo{journal}{Journal of Mathematical imaging and vision}
  \bibinfo{volume}{20} (\bibinfo{year}{2004}) \bibinfo{pages}{89--97}.
\bibitem[{Jenatton et~al.(2012)Jenatton, Gramfort, Michel, Obozinski, Eger,
  Bach, and Thirion}]{jenatton2012multiscale}
\bibinfo{author}{R.~Jenatton}, \bibinfo{author}{A.~Gramfort},
  \bibinfo{author}{V.~Michel}, \bibinfo{author}{G.~Obozinski},
  \bibinfo{author}{E.~Eger}, \bibinfo{author}{F.~Bach},
  \bibinfo{author}{B.~Thirion},
\newblock \bibinfo{title}{Multiscale mining of fmri data with hierarchical
  structured sparsity},
\newblock \bibinfo{journal}{SIAM Journal on Imaging Sciences}
  \bibinfo{volume}{5} (\bibinfo{year}{2012}) \bibinfo{pages}{835--856}.
\bibitem[{Watanabe et~al.(2014)Watanabe, Kessler, Scott, Angstadt, and
  Sripada}]{watanabe2014disease}
\bibinfo{author}{T.~Watanabe}, \bibinfo{author}{D.~Kessler},
  \bibinfo{author}{C.~Scott}, \bibinfo{author}{M.~Angstadt},
  \bibinfo{author}{C.~Sripada},
\newblock \bibinfo{title}{Disease prediction based on functional connectomes
  using a scalable and spatially-informed support vector machine},
\newblock \bibinfo{journal}{Neuroimage} \bibinfo{volume}{96}
  (\bibinfo{year}{2014}) \bibinfo{pages}{183--202}.
\bibitem[{Ehrgott(2005)}]{ehrgott2005multicriteria}
\bibinfo{author}{M.~Ehrgott}, \bibinfo{title}{Multicriteria optimization},
  volume \bibinfo{volume}{491}, \bibinfo{publisher}{Springer Science \&
  Business Media}, \bibinfo{year}{2005}.
\bibitem[{Zhang et~al.(2019)Zhang, He, Sra, and Jadbabaie}]{gradientclipping}
\bibinfo{author}{J.~Zhang}, \bibinfo{author}{T.~He}, \bibinfo{author}{S.~Sra},
  \bibinfo{author}{A.~Jadbabaie},
\newblock \bibinfo{title}{Why gradient clipping accelerates training: A
  theoretical justification for adaptivity},
\newblock \bibinfo{journal}{arXiv preprint arXiv:1905.11881}
  (\bibinfo{year}{2019}).
\bibitem[{Keskar et~al.(2016)Keskar, Mudigere, Nocedal, Smelyanskiy, and
  Tang}]{keskar2016large}
\bibinfo{author}{N.~S. Keskar}, \bibinfo{author}{D.~Mudigere},
  \bibinfo{author}{J.~Nocedal}, \bibinfo{author}{M.~Smelyanskiy},
  \bibinfo{author}{P.~T.~P. Tang},
\newblock \bibinfo{title}{On large-batch training for deep learning:
  Generalization gap and sharp minima},
\newblock \bibinfo{journal}{arXiv preprint arXiv:1609.04836}
  (\bibinfo{year}{2016}).
\bibitem[{Davis and Goadrich(2006)}]{davis2006relationship}
\bibinfo{author}{J.~Davis}, \bibinfo{author}{M.~Goadrich},
\newblock \bibinfo{title}{The relationship between precision-recall and roc
  curves},
\newblock in: \bibinfo{booktitle}{Proceedings of the 23rd international
  conference on Machine learning}, \bibinfo{year}{2006}, pp.
  \bibinfo{pages}{233--240}.
\bibitem[{Paszke et~al.(2019)Paszke, Gross, Massa, Lerer, Bradbury, Chanan,
  Killeen, Lin, Gimelshein, Antiga, Desmaison, Kopf, Yang, DeVito, Raison,
  Tejani, Chilamkurthy, Steiner, Fang, Bai, and Chintala}]{NEURIPS2019_9015}
\bibinfo{author}{A.~Paszke}, \bibinfo{author}{S.~Gross},
  \bibinfo{author}{F.~Massa}, \bibinfo{author}{A.~Lerer},
  \bibinfo{author}{J.~Bradbury}, \bibinfo{author}{G.~Chanan},
  \bibinfo{author}{T.~Killeen}, \bibinfo{author}{Z.~Lin},
  \bibinfo{author}{N.~Gimelshein}, \bibinfo{author}{L.~Antiga},
  \bibinfo{author}{A.~Desmaison}, \bibinfo{author}{A.~Kopf},
  \bibinfo{author}{E.~Yang}, \bibinfo{author}{Z.~DeVito},
  \bibinfo{author}{M.~Raison}, \bibinfo{author}{A.~Tejani},
  \bibinfo{author}{S.~Chilamkurthy}, \bibinfo{author}{B.~Steiner},
  \bibinfo{author}{L.~Fang}, \bibinfo{author}{J.~Bai},
  \bibinfo{author}{S.~Chintala},
\newblock \bibinfo{title}{Pytorch: An imperative style, high-performance deep
  learning library},
\newblock in: \bibinfo{booktitle}{NeurIPS}, \bibinfo{publisher}{Curran
  Associates, Inc.}, \bibinfo{year}{2019}, pp. \bibinfo{pages}{8024--8035}.
\bibitem[{Pedregosa et~al.(2011)Pedregosa, Varoquaux, Gramfort, Michel,
  Thirion, Grisel, Blondel, Prettenhofer, Weiss, Dubourg
  et~al.}]{pedregosa2011scikit}
\bibinfo{author}{F.~Pedregosa}, \bibinfo{author}{G.~Varoquaux},
  \bibinfo{author}{A.~Gramfort}, \bibinfo{author}{V.~Michel},
  \bibinfo{author}{B.~Thirion}, \bibinfo{author}{O.~Grisel},
  \bibinfo{author}{M.~Blondel}, \bibinfo{author}{P.~Prettenhofer},
  \bibinfo{author}{R.~Weiss}, \bibinfo{author}{V.~Dubourg}, et~al.,
\newblock \bibinfo{title}{Scikit-learn: Machine learning in python},
\newblock \bibinfo{journal}{the Journal of machine Learning research}
  \bibinfo{volume}{12} (\bibinfo{year}{2011}) \bibinfo{pages}{2825--2830}.
\bibitem[{McKinney et~al.(2011)}]{mckinney2011pandas}
\bibinfo{author}{W.~McKinney}, et~al.,
\newblock \bibinfo{title}{pandas: a foundational python library for data
  analysis and statistics},
\newblock \bibinfo{journal}{Python for High Performance and Scientific
  Computing} \bibinfo{volume}{14} (\bibinfo{year}{2011}).
\bibitem[{Waskom(2021)}]{waskom2021seaborn}
\bibinfo{author}{M.~Waskom},
\newblock \bibinfo{title}{Seaborn: statistical data visualization},
\newblock \bibinfo{journal}{Journal of Open Source Software}
  \bibinfo{volume}{6} (\bibinfo{year}{2021}) \bibinfo{pages}{3021}.

\end{thebibliography}
\clearpage

\appendix
\section{Convergence of our training method}

\begin{figure}[H]
\begin{minipage}{165mm}
\begin{subfigure}{0.9\textwidth}
  \centering
  \includegraphics[scale=0.4]{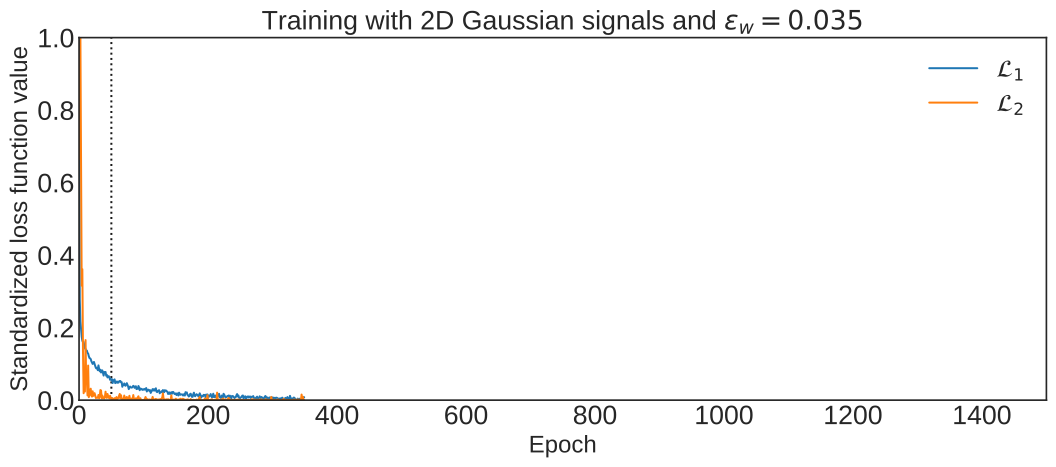}
  \label{fig:sfig1}
\end{subfigure}%

\begin{subfigure}{0.9\textwidth}
  \centering
  \includegraphics[scale=0.4]{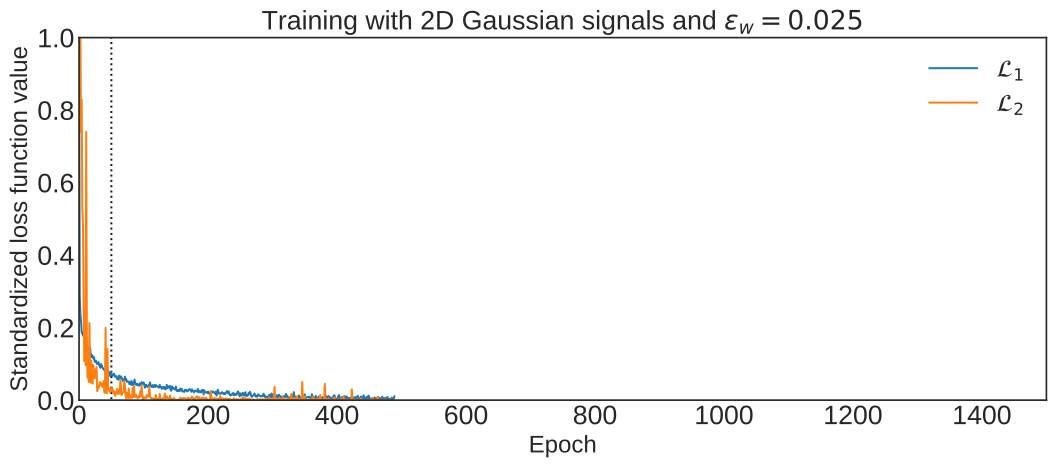}
  \label{fig:sfig2}
\end{subfigure}

\begin{subfigure}{0.9\textwidth}
  \centering
  \includegraphics[scale=0.4]{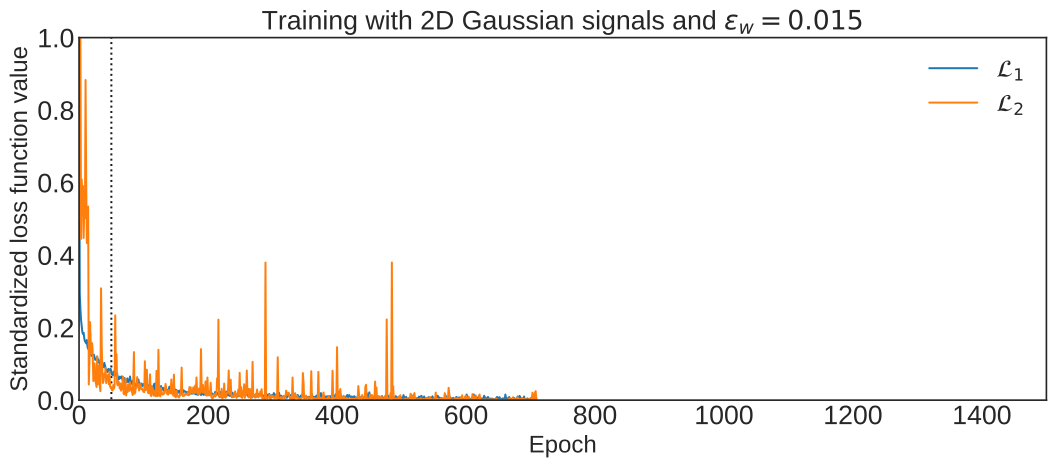}
  \label{fig:sfig6}
\end{subfigure}%

\begin{subfigure}{0.9\textwidth}
  \centering
  \includegraphics[scale=0.4]{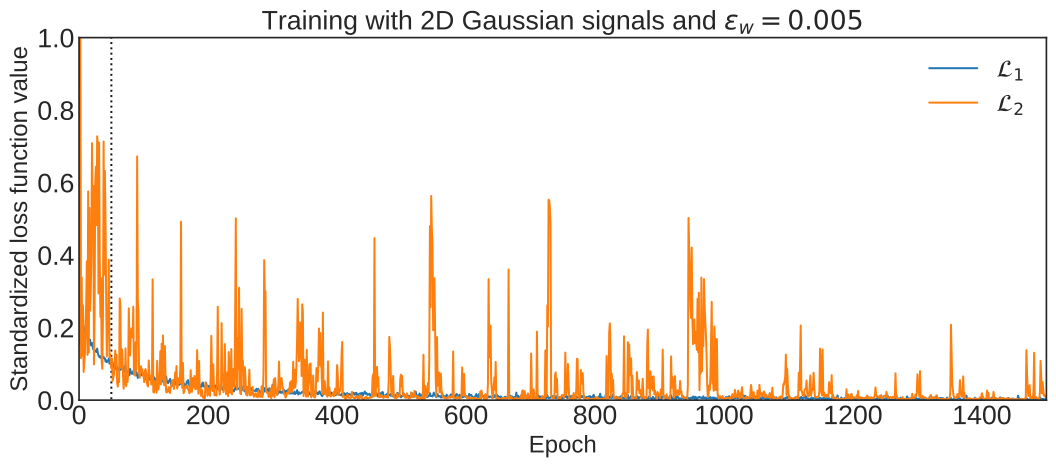}
  \label{fig:sfig3}
\end{subfigure}
\caption{Sample of convergence behavior during the training process. The dotted lines represent the end of the initial phase where the masks are defined. An early stopping criterion with patience equal to 10 is applied. Let $\mathcal{L}_1$ be the primary loss for classification and $\mathcal{L}_2$ represents the secondary loss for regularization. Note that the local minimum obtained for each training is different; hence, a faster convergence/stopping does not directly imply a better hyperparameter selection $\epsilon_w$. In fact, there is a trade-off between $\mathcal{L}_1$ and $\mathcal{L}_2$ which is managed by  $\epsilon_w$. } 
\label{convergence}
\end{minipage}
\end{figure}

\color{black}
\section{List of features}

\begin{table}[H]
\caption{List of feature sets used to train the MLP classifier. $n$ represents the number of features estimated by each set. They were estimated using \cite{nun2015fats}.}
\centering
\color{black}
\begin{tabular}{|l|c|l|c|}
\hline

Feature name &  Number of features & Feature name &  Number of features \\
\hline
  PeriodLS           &  1   & FluxPercentilRationMid &     5 \\
  AndersonDarling    &    1  & Freq$_s$\_harmonics &         21 \\
  Amplitude           &    1 &  Gskew   &   1 \\
  Beyond1Std          &    1 &  LinearTrend   &       1 \\
  CAR                 &    3      &  MaxSlope &      1 \\
  MedianBRP            &    1        &  Mean &       1 \\
  PairSlopeTrend      &    1        & MeanVariance &       1 \\
  PercentAmplitude    &    1        & MedianAdsDev &       1 \\
  PercentDiferenceFluxPercentile              & 1       &  Period\_fit      & 1 \\
  Psi              & 2       &  Q31 &       1 \\
  Rcs              & 1       &  Skew &       1 \\
  SlottedA\_length             &  1       & Small\_Kurtosis&       1 \\
  Std              &      1  & Stentson &      2\\
 StructureFunction\_index            &   3     & Eta       & 1 \\
 Con         &     1     & Autocor\_length       &  1 \\
   \hline
\end{tabular}

\label{features}
\end{table}
\color{black}





\end{document}